\documentclass{aa}  

\usepackage{graphicx}
\usepackage{txfonts}
\usepackage{natbib}
\usepackage{color}
\bibpunct{(}{)}{;}{a}{}{,} 

\begin{document} 



\title{\textit{Gaia} DR2 Proper Motions of Dwarf Galaxies within 420 kpc:} 

\subtitle{Orbits, Milky Way Mass, Tidal Influences, Planar Alignments, and Group Infall}

   \author{T. K. Fritz
          \inst{1,2}
          \and
          G. Battaglia\inst{1,2}
          \and M. S. Pawlowski\inst{3}\thanks{Hubble fellow}
          \and N. Kallivayalil\inst{4}
          \and R. van der Marel\inst{5,6}
          \and  S. T. Sohn \inst{5}
          \and C. Brook \inst{1,2}\
          \and G. Besla \inst{7}
          }

   \institute{Instituto de Astrofisica de Canarias, calle Via Lactea s/n, E-38205 La Laguna, Tenerife, Spain\\
              \email{tfritz@iac.es}
         \and
         Universidad de La Laguna, Dpto. Astrofisica, E-38206 La Laguna, Tenerife, Spain
       \and Department of Physics and Astronomy, University of California, Irvine, CA 92697, USA 
       \and Department of Astronomy, University of Virginia, Charlottesville, 530 McCormick Road, VA 22904-4325, USA
       \and Space Telescope Science Institute, 3700 San Martin Drive, Baltimore, MD 21218, USA
\and Center for Astrophysical Sciences, Department of Physics \& Astronomy, Johns Hopkins University, Baltimore, MD 21218, USA
       \and Steward Observatory, University of Arizona, 933 North Cherry Avenue, Tucson, AZ 85721, USA
             }

   \date{}

 
  \abstract{
  {A proper understanding of the Milky Way (MW) dwarf galaxies
in a cosmological context requires knowledge of their 3D velocities
and orbits. However, proper motion (PM) measurements have generally
been of limited accuracy and available only for more massive dwarfs.}
{We therefore present a new study of the kinematics of the MW
dwarf galaxies.}
{We use the \textit{Gaia} DR2 for those dwarfs that have
  been spectroscopically observed in the literature}.
{We derive systemic PMs for 39 galaxies  and galaxy candidates out to 420\,kpc, and
generally find good consistency for the subset with measurements
available from other studies. We derive the implied Galactocentric
velocities, and calculate orbits in canonical MW halo potentials of
"low" ($0.8 \times 10^{12} M_\odot$) and "high" mass ($1.6 \times
10^{12} M_\odot$).}
{Comparison of the distributions of orbital apocenters
and 3D velocities to the halo virial radius and escape velocity,
respectively, suggests that the satellite kinematics are best explained
in the high-mass halo. 
Tuc~III, Crater~II, and additional
candidates have orbital pericenters small enough to imply significant
tidal influences. Relevant to the missing satellite problem, the fact
that fewer galaxies are observed to be near apocenter than near
pericenter implies that there must be a population of distant dwarf
galaxies yet to be discovered. 
Of the 39 dwarfs: 12 have orbital
poles that do not align with the MW
plane of satellites (given reasonable assumptions about its intrinsic
thickness); 10 have insufficient PM accuracy to establish whether
they align; and 17 satellites align, of which 11 are co-orbiting and
(somewhat surprisingly, in view of prior knowledge) 6 are counter-orbiting.
Group infall might have contributed to this, but no definitive
association is found for the members of the Crater-Leo group.
}

     }


\titlerunning{\textit{Gaia} DR2 Proper Motions and Orbits of Dwarf Galaxies within 420 kpc}
\authorrunning{T. Fritz et al.}   

   \maketitle
%

\section{Introduction}

The determination of the orbital properties of the dwarf galaxies surrounding
the Milky Way (MW) is a crucial step for unraveling the formation and
evolutionary paths of these galaxies: did the
dwarf spheroidals (dSphs) and ultra-faint dwarfs (UFDs) 
lose their gas due to internal stellar feedback or due to a combination
of internal and external effects, such as UV-background heating and/or
tidal and ram-pressure stripping from the large Local Group
spirals \citep[e.g.][]{Mayer_06, Bermejo-Climent_2018, Revaz_18}?
Are the metallicity gradients observed for several of the Milky Way 
``classical'' dwarf spheroidal galaxies \citep[e.g.][]{Tolstoy_04, Battaglia_06, Battaglia_11} an intrinsic property
of these systems or could interactions with
the Milky Way have had a role in setting their
presence or erasing them in some cases 
\citep{Sales_10}? 

From a theoretical perspective,
depending on the orbital and internal characteristics of the object
(e.g. in terms of the mass density profile of its dark matter halo) it has
been argued that such interactions could potentially  
transform initially discy/rotating dwarfs in spheroidal/pressure-supported systems \citep[e.g.][]{Mayer_01,Mayer_06,Kazantzidis_17}; 
however, the
existence of very isolated dwarf galaxies, or those likely on their first approach towards the MW, having a 
spheroidal morphology and being in the process of losing/exhausting their gas would suggest otherwise
(e.g. VV~124, Leo~T, Phoenix~I).  Nonetheless, signs of tidal disturbance
have been detected in the spatial distribution of the stellar component of
some classical dwarf galaxies and ultra-faint systems
\citep[e.g. Carina~I, Hercules~I, Bo\"{o}tes~I, see e.g.][]{Battaglia_12, Roderick_15, Roderick_16} and it is still debated
whether the presence of tidally unbound stars in spectroscopic samples can
alter inferences of the dark matter (DM) distribution in these galaxies obtained from
the observed line-of-sight velocity dispersion, in particular in the outer parts \citep[e.g.][]{Lokas_08, Lokas_09}. 

Other interesting aspects that the orbital properties allow to explore are  
whether some of the MW satellites were accreted to the MW as part 
of a group of dwarf galaxies as suggested for example by the discovery of possible
satellites of the LMC \citep{Koposov_15a,Bechtol_15,Martin_15} as well as if they are preferentially
distributed on a planar structure \citep{Pawlowski_12}.

These are only
a few of the questions that can potentially be addressed by knowledge
of the orbital properties of the MW dwarf galaxies. The very first step
towards this goal is the determination of the 3D bulk motion of these systems.
While determinations of the systemic heliocentric line-of-sight (l.o.s.) velocity are available in the
literature for all of the MW classical dwarf spheroidal galaxies and most
of the ultra-faint systems discovered so far, astrometric
proper motions have became available for the full set of MW classical dwarf
spheroidals only very recently and no astrometric proper motion had been
available for any of the ultra-faint dwarf galaxies, apart from
Segue~1 \citep{Fritz_18}. 

The second release of data from the \textit{Gaia} mission \citep{Brown_18} has started a
revolution in this respect. \citet{Helmi_18} demonstrated the power of the
\textit{Gaia} DR2 data for the study of the kinematics of stellar systems around the
Milky Way using 75 globular clusters, the Magellanic Clouds, the 9 classical
MW dwarf spheroidal galaxies and the Bo\"{o}tes~I UFD.
In several cases the precisions are exquisite, in others they are comparable to
what can be achieved with HST, but with the clear advantage of being in an
absolute reference frame.  

\textit{Gaia} DR2 has opened the door as well to the 
determinations of the systemic proper motion of dozens of the
other dwarf galaxies surrounding the Milky Way. In this work,
we obtain systemic proper motions for most of the systems within 420\,kpc from the MW. The paper is 
structured as follows: in Sect.~2 we describe the data-sets used for the analysis and in Sect.~3 the methodology 
adopted to determine systemic proper motions; Sect.~4 presents the determination of space velocities and orbital properties 
of the objects in the sample; we discuss the main results in Sect.~5 and present conclusions and summary in Sect.~6.

\section{Data}

Our sample of galaxies consists of the vast majority of galaxies and galaxy candidates within 420\,kpc of the Milky~Way. We omitted: the Magellanic Clouds, because they have already well established motions and their measurable internal motions complicate measurements; 
Leo~T, because there is only a handful of  stars with spectroscopic data that have matches in \textit{Gaia}, and these are very faint, and therefore with large proper motion errors; and Bootes~III, due to 
its unclear nature as a galaxy or stream \citep{Grillmair_09}.  We were inclusive of objects whose nature as globular cluster or as dwarf galaxy is still under debate, such as for example Crater~I \citep{Kirby_15,Voggel_16}. 

Our work is mainly based on the \textit{Gaia} \citep{2016A&A...595A...1G} data release 2 (GDR2) \citep{Brown_18}. Furthermore, we use literature spectroscopic data sets of dwarf galaxies to ease identification of members (see Table~\ref{KapSou}). 
 
Because we rely on spectroscopic data, some satellites are omitted from our sample; for instance, we drop Pegasus~III because of the lack of \textit{Gaia} DR2 matches for the spectroscopic members  
in \citet{Kim_16a}. Some other systems are so faint and/or distant (for example, Indus~II) that calculations show that no member stars are expected above the \textit{Gaia} DR2 magnitude limit. Others, like Sagittarius~II and Pictor~II, are in principle bright enough, but no spectroscopic data-set has been published yet.

      \begin{table*}
      \caption[]{Sample of objects. Column 1 lists the object name; col. 2 the number of spectroscopic members; col. 3 gives the number of spectroscopic members that have a match within 1'' in \textit{Gaia} DR2 and col. 4 is as before 
but that have kinematic information in \textit{Gaia} DR2; col. 5 lists the number of spectroscopic members that passed 
our additional membership criteria (see text); in col.6 we provide the source of the spectroscopic catalogs. 
In column 6 we also mark  the cases for which we derived the spectroscopic membership probability ourselves, next to the catalog where this was done: "Yv", when only l.o.s. velocities were used, "Yvg" or "YvMg" when also the information on the star's log$g$ or the near-IR Mg~I line at 8806.8\AA\, were taken into account. In column 7 and 8 we list the distance modulus adopted and its source, respectively; note that we always add an error of 0.1\,mag in quadrature to the error in distance modulus  listed here, to safeguard against underestimated systematic errors. }
         \label{KapSou}
     $$
         \begin{array}{p{0.15\linewidth}lllllll}
            \hline
            name & \mathrm{spec~mem} & \mathrm{in}~Gaia & \mathrm{kinematic} & \mathrm{final~sample} & \mathrm{spec~source} & \mathrm{dm} & \mathrm{dm~source}  \\
            \hline
 Aquarius II & 9 & 2 & 2 & 2 & 1 & 20.16\pm0.07 & 1 \\
 Bo\"{o}tes I & 78 & 45 & 38 & 38 & 2, 30Yv &19.11\pm0.08 & 32 \\
 Bo\"{o}tes II & 5 & 4 & 4 & 4 & 3 & 18.11\pm0.06 & 33\\
 CanVen I & 237 & 69 & 57 & 57 & 2, 4 &21.62 \pm 0.05 & 34\\
 CanVen II & 25 & 13 & 11 & 11 & 4 & 21.02 \pm 0.06 & 35\\
 Carina I & 780 & 772 & 693 & 693 & 5 & 20.08\pm  0.08 & 57, 58 \\
 Carina II & 18 & 18 & 18 & 18 & 6 & 17.79\pm0.05 & 36 \\
 Carina III & 4 & 4 & 4 & 4 & 6 &17.22\pm0.10 & 36\\
 Coma Berenices I & 59 & 18 & 18 & 17 & 4 & 18.13\pm 0.08 & 37\\
 Crater I & 36 & 10 & 10 & 10 & 7, 8 & 20.81\pm0.04 & 38\\
 Crater II & 59 & 59 & 58 & 58 & 9, Yvg & 20.25\pm0.10 & 39\\
  Draco I & 496 & 495 & 440 & 436 & 10, Yvg & 19.49\pm 0.17 & 59, 60 \\
 Draco II & 9 & 6 & 6 & 6 & 11 & 16.66\pm 0.04 & 40\\ 
  Eridanus II & 28 & 13 & 12 & 12 & 12 & 22.8\pm0.1 & 41 \\
 Fornax I & 2906 & 2891 & 2547 & 2527 & 5, 13YvMg & 20.72\pm0.04 & 61\\
  Grus I & 8 & 6 & 6 & 6 & 14 & 20.4\pm0.2 & 44\\
 Hercules I & 47 & 26 & 22 & 22 & 4, 15 & 20.64\pm0.14 & 42, 43 \\
 Horologium I & 5 & 5 & 4 & 4 & 16 &19.6\pm0.2 & 44, 45 \\
 Hydra II & 13 & 6 & 6 & 6 & 7 & 20.89\pm 0.12 & 46 \\
 Hydrus I & 33 & 33 & 32 & 30 & 29 & 17.20 \pm0.04 & 29  \\ 
 Leo I & 328 & 299 & 241 & 241 & 17, Yv & 22.15\pm 0.1 & 62  \\
 Leo II & 246 & 142 & 131 & 131 & 18, 31 & 21.76\pm 0.13 & 63, 64 \\
 Leo IV & 18 & 5 & 3 & 3 & 4 & 20.94 \pm 0.07 & 47\\
 Leo V & 8 & 5 & 5 & 5 & 19 &21.19 \pm 0.06 & 48  \\
 Phoenix I & 194 & 83 & 71 & 71 & 20 & 23.11\pm 0.1 & 66 \\
 Pisces II & 7 & 2 & 2 & 2 & 7 & 21.31\pm0.18 & 49 \\
 Reticulum II & 28 & 28 & 27 & 27 & 21 &17.5\pm0.1 & 50 \\
 Sagittarius I & 151 & 151 & 124 & 96 & \mathrm{APOGEE}, Yvg & 17.13\pm0.11 & 65 \\
 Sculptor I & 1661 & 1652 & 1483 & 1468 & 5, 13YvMg & 19.64\pm 0.13 &  67, 68\\
 Segue 1 & 71 & 15 & 14 & 13 & 22 & 16.8\pm 0.2 & 51 \\
 Segue 2 & 26 & 13 & 10 & 10 & 23 &17.8\pm0.18 & 52 \\ 
 Sextans I & 549 & 392 & 328 & 325 &  24  & 19.67\pm 0.15 & 68 \\
 Triangulum II & 13 & 6 & 5 & 5 & 25& 17.27\pm0.1 & 53  \\
 Tucana II & 27 & 19 & 19 & 19 & 14 & 18.8\pm 0.2 & 44, 45\\ 
 Tucana III & 50 & 42 & 40 & 39 & 26, 27 & 16.8\pm 0.1 & 50\\ 
 Ursa Major I & 40 & 29 & 23 & 23 & 2, 4 & 19.94\pm 0.13 & 54\\
 Ursa Major II & 28 & 17 & 15 & 15 & 2, 4 & 17.70\pm 0.13 & 55\\
 Ursa Minor I & 212 & 152 & 137 & 137 & 28 & 19.40\pm 0.11 & 70, 71\\
 Willman 1 & 14 & 8 & 7 & 7 & 2 & 17.90\pm0.40 & 56 \\
            \noalign{\smallskip}
            \hline
         \end{array}
  $$   
  \tablebib{(1)~\citet{Torrealba_16a};
(2) \citet{Martin_07}; (3) \citet{Koch_09}; (4) \citet{Simon_07};
(5) \citet{Walker_09_clas}; (6) \citet{Li_18a}; (7) \citet{Kirby_15};
(8) \citet{Voggel_16}; (9) \citet{Caldwell_17}; (10) \citet{Walker_15};
(11) \citet{Martin_16}; (12) \citet{Li_17}; (13) \citet[][and references therein]{BattagliaSt_12}; (14) \citet{Walker_16}; (15) \citet{Aden_09}; (16) \citet{Koposov_15b}; (17) \citet{Mateo_08}; (18)  \citet{Spencer_17}; (19) \citet{Walker_09}; (20) \citet{Kacharov_17}; (21) \citet{Simon_15}; (22) \citet{Simon_11}; (23) \citet{Kirby_13}; (24) \citet{Cicuendez_18}; (25) \citet{Kirby_17}; (26) \citet{Simon_17}; (27) \citet{Li_18b}; (28)  \citet{Kirby_10}; (29) \citet{Koposov_18}; (30) \citet{Koposov_11}; (31) \citet{Koch_07}; (32) \citet{DallOra_06}; (33) \citet{Walsh_08}; (34) \citet{Kuehn_08}; (35) \citet{Greco_08}; (36) \citet{Torrealba_18}; (37) \citet{Musella_09}; (38) \citet{Weisz_16}; (39) \citet{Joo_18}; (40) \citet{Longeard_18}; (41) \citet{Crnovic_16}; (42) \citet{Musella_12}; (43) \citet{Garling_18}; (44) \citet{Koposov_15a}; (45) \citet{Bechtol_15}; (46)\citet{Vivas_16}; (47) \citet{Moretti_09}; (48) \citet{Medina_17}; (49) \citet{Sand_12}; (50) \citet{Mutlu_18}; (51) \citet{Belokurov_07}; (52) \citet{Boetcher_13}; (53) \citet{Carlin_17}; (54) \citet{Garofalo_13}; (55) \citet{DallOra_12}; (56)  \citet{Willman_06}, (57) \citet{Coppola_15}; (58) \citet{Vivas_13}; (59) \citet{Bonanos_04}; (60) \citet{Kinemuchi_08}; (61) \citet{Rizzi_07}; (62) \citet{Stetson_14}; (63) \citet{Bellazzini_05}; (64) \citet{Gullieuszik_08}; (65) \citet{Hamanowicz_16}; (66) \citet{Holtzman_00}; (67) \citet{Martinez_16}; (68) \citet{Pietrzynski_08}; (69) \citet{Mateo_95}; (70) \citet{Carrera_02}; (71) \citet{Bellazzini_02} .
}
   \end{table*}

\section{Proper motions}

\subsection{Method}

 As a first step, we perform a non-restrictive selection based on the information from spectroscopic measurements, mainly aimed at excluding obvious contaminants. This is made by retaining stars with at least 40\% probability of being members to a given galaxy according to the spectroscopic analysis. Except for a couple of cases, membership probabilities (be it a  binary or continuous probability) are available  from the literature. For those cases when these are  
not provided in the source paper, we calculate them ourselves (see notes in Tab.~\ref{KapSou}) using the heliocentric velocities, $v_\mathrm{LOS}$, and, when available, also gravity indicators: essentially we assign probabilities based on the likelihood of star to be a giant and 
have a velocity similar to the systemic velocity of the galaxy, see Appendix~\ref{ap_det_galaxies} for these and other details on the target selection.

 This first, broad selection of probable members is later refined by including information from \textit{Gaia} DR2. We first check  
whether stars with spectroscopic information have a match within 1" in \textit{Gaia} DR2  and have also \textit{Gaia} DR2 kinematic 
information. We then require  the stars to have a parallax deviating less than 2$\sigma$\footnote{ Throughout this work, unless stated otherwise, by $\sigma$ we mean the measurement error on a given quantity.} from the value  expected given the distance 
to the galaxy, and tangential velocities (in the right ascension and declination direction) 
less  than twice the corresponding measurement error from an approximation of the escape speed at the distance of the 
object. Our estimate of the local escape speed corresponds approximately to that
expected for a Navarro, Frenk and White (NFW) dark matter (DM) halo \citep{NFW_96} with virial mass 1.2$\times 10^{12}$M$_{\odot}$. This is a generous selection, in which objects with fairly extreme velocities such as Leo~I, Bo\"{o}tes~III and the Magellanic Clouds are required to be bound to the MW. 
After converting the escape speed to mas/yr at the distance of the object, we also add the expected 'proper motion' due to the reflex motion of the Sun given the position and distance of the satellite.
The escape speed criteria is applied in both dimensions (right ascension and declination) separately. All these criteria are relatively inclusive, but, because we use several of them, the sample is still expected to be rather clean. 

Finally, we apply an outlier rejection criteria, excluding those stars whose proper motion deviates by more than 3 $\sigma$ in at least one proper motion dimension, where in this case $\sigma$ is the measurement error in the individual proper motion. 

Overall not many stars are excluded from the initial samples, see Table~\ref{KapSou}, especially for the fainter galaxies. 
We also explored changing our  spectroscopic probability cut. Even when we used a spectroscopic membership probability $>$75\%, the change in the resulting values of the systemic proper motion was at most, but typically  much less than, 0.5\,$\sigma$. We list the few exceptions in the appendix. We note that for most of the faint galaxies the membership  probability provided in the source papers is binary and thus always the same stars are selected.
A few objects have only two likely members with kinematic information in \textit{Gaia} DR2 and passing our criteria 
(see Table~\ref{KapSou}). The final step of outlier rejection cannot therefore be applied on these samples containing only two stars. We inspected the motions of these cases with greater care; most of them agree well with each other, and we use them in the following w/o distinction, see appendix for details. 
Especially for objects for which only an handful of spectroscopic members are available, we also double checked that \textit{Gaia} DR2 stars 
within one half-light radius from the center of these systems, but without spectroscopic measurements, clump close to 
the member stars in proper motion space. For simplicity we omitted to add these others stars to our sample.  

After we have selected which stars are members, we calculate the average proper motion of the object in
right ascension ($\mu_{\alpha^*})$ and declination ($\mu_{\delta})$ and the average correlation coefficient between $\mu_{\alpha*}$ and $\mu_\delta$ by taking an error weighted average of all the member stars. These values are given in Table~\ref{KapSou2}. In this table we also define the abbreviations for satellite names used in this paper. For the conversion to velocities and for our analysis we also add a systematic error to the proper motions errors in both dimensions, which we assume to be uncorrelated between $\alpha$ and $\delta$. For this error, we use a value of 0.035 mas/yr for galaxies which cover at least 0.2\,degree \citep[see][]{Helmi_18}. For smaller galaxies we use an error of 0.066 mas/yr \citep[Table~4 of][]{Lindegren_18} for galaxies of zero size and we interpolate linearly between these two cases. We list the systematic error adopted in Table~\ref{KapSou2}. Even though the smallest galaxies tend to have the largest systematic errors, these are  always smaller than the statistical errors. On the other hand, for most of the classical dwarf galaxies, the systematic error dominates.

The obtained proper motions are of high quality as shown by the comparison with independent measurements
(see Sect.~\ref{comp-meas}) and by the reduced $\chi^2$ distribution between individual proper motion measurements and sample averages, which  
follows what is expected for samples of the sizes considered here.  Note that while in the inner halo there are still many halo stars in streams and in the
smooth halo component, we expect the contamination from halo stars to be much lower at large Galactocentric distances.
Thus the motions of distant satellites will be more reliable in that respect.

\subsection{Comparison with other measurements}
\label{comp-meas}

\textit{Gaia} DR2 proper motions are known to have systematic uncertainties
of the order $0.035$ to 0.066 mas/yr \citep{Helmi_18,Lindegren_18}.
Since some of the closer galaxies in our sample have measurement uncertainties
smaller or comparable to this, it is important to compare our
\textit{Gaia} DR2 proper motions with those measured using independent instruments
and methods. \citet{Helmi_18} compared their
\textit{Gaia} DR2 proper motion measurements of 9 classical dSphs with those
reported in the literature. They found that in general, the \textit{Gaia} DR2
measurements are consistent with previous ones (based on
ground-based and HST data) at 2\,$\sigma$ level. Moreover, when
compared to only HST-based measurements, the agreement becomes even
better, especially when systematic uncertainties are considered.
Our proper motions presented in Table~\ref{KapSou2} are consistent with the
results by \citet{Helmi_18} within the quoted $1\sigma$ uncertainties
for 6 out of the 9 classical dSphs. The exceptions are large galaxies on the sky,  where systematic errors and real physical differences might be more important, see Appendix~\ref{ap_det_galaxies} for the details.  
Thus, our results for classical dSphs are generally in
good agreement with \citet{Helmi_18}, so we are able to make the
same assessments as in their paper. 

For the UFDs, the only object for which a comparison to non-\textit{Gaia}-based proper motions can be made is  Segue~1; however, due to a clearly lower precision of this measurement, it is not so useful to judge the goodness of our results (Appendix~\ref{ap_det_galaxies}). For the other UFDs, there are other \textit{Gaia}-based  measurements available \citep{Simon_18,Massari_18,Kallivayalil_18,Pace_18}, which can be used to check the results against possible sources of errors such as selection of members stars, but not for systematic errors. We compare in detail to these works in appendix~\ref{ap_det_galaxies}. In most cases we agree better than 1\,$\sigma$ , but such a good agreement is to be expected, because some of these  determinations are not truly independent since they use the same spectroscopic data-sets for member selection or as a starting guess. A difference larger than 2.5\,$\sigma$ occurs for Crater~II and Tucana~III compared to \citet{Kallivayalil_18}, which is possibly due to a non-optimal treatment of the background contamination. Our measurement disagrees with both the other two existing measurements for Segue~2 (which also disagree with each other). This galaxy is a difficult case due to its faintness and a large foreground contamination, which prevents a safe identification of true members.

     \begin{table*}
      \caption[]{Distance and kinematic properties of objects in our sample. Column 1 lists the abbreviated satellite name (the satellites are in the same order as in Table~1), col. 2 its 
Galactocentric distance, col.3 and 4 give the proper motion we obtain, followed by the statistical error and the second the systematic error, col. 5 the correlation coefficient between $\mu_{\alpha,*}$ and $\mu_{\delta}$, col. 6 the heliocentric systemic $V_\mathrm{LOS}$ (the source of which is typically the same as the spectroscopic catalogs we adopt); 
col. 7, 8 and 9 list the Galactocentric 3D, radial and tangential velocity, respectively  Galactocentric velocity. We warn the reader that, 
since in some cases the errors are large and due to the existing correlation between the two proper motion directions, 
for detailed calculations it is best to start from the measured quantities (i.e. the proper motions).}
         \label{KapSou2}
       $
         \begin{array}{p{0.06\linewidth}llllllll}
            \hline
            satellite & d_{GC}& \mu_{\alpha^*} & \mu_\delta & C_{\mu_{\alpha*},\mu_\delta} &  V_\mathrm{LOS}  &V_\mathrm{3D} & V_\mathrm{rad} &  V_\mathrm{tan}\\
             & [kpc]& [mas/yr] &  [mas/yr] & & [km/s] & [km/s] & [km/s] & [km/s]\\            
            \hline
AquII & 105 & -0.252\pm0.526\pm0.063 & 0.011\pm0.448\pm0.063 & 0.131 & -71.1\pm2.5 & 250^{+241}_{-164} & 49\pm8 & 244^{+242}_{-174} \\
BooI & 64 & -0.554\pm0.092\pm0.035 & -1.111\pm0.068\pm0.035 & 0.163 & 99\pm2.1 & 192^{+27}_{-25} & 94\pm2 & 167^{+32}_{-31} \\
BooII & 39 & -2.686\pm0.389\pm0.056 & -0.53\pm0.287\pm0.056 & -0.186 & -117\pm5.2 & 383^{+76}_{-68} & -54\pm9 & 379^{+79}_{-70} \\
CVenI & 211 & -0.159\pm0.094\pm0.035 & -0.067\pm0.054\pm0.035 & 0.105 & 30.9\pm0.6 & 124^{+68}_{-38} & 82\pm2 & 94^{+79}_{-63} \\
CVenII & 161 & -0.342\pm0.232\pm0.056 & -0.473\pm0.169\pm0.056 & -0.006 & -128.9\pm1.2 & 183^{+150}_{-77} & -93\pm4 & 157^{+163}_{-108} \\
CarI & 105 & 0.485\pm0.017\pm0.035 & 0.131\pm0.016\pm0.035 & 0.083 & 229.1\pm0.1 & 163^{+21}_{-22} & 2\pm2 & 163^{+21}_{-22} \\
CarII & 37 & 1.867\pm0.078\pm0.035 & 0.082\pm0.072\pm0.035 & -0.008 & 477.2\pm1.2 & 355^{+16}_{-14} & 203\pm3 & 291^{+19}_{-18} \\
CarIII & 29 & 3.046\pm0.119\pm0.057 & 1.565\pm0.135\pm0.057 & 0.066 & 284.6\pm3.4 & 388^{+33}_{-30} & 46\pm4 & 385^{+33}_{-30} \\
CBerI & 43 & 0.471\pm0.108\pm0.035 & -1.716\pm0.104\pm0.035 & -0.427 & 98.1\pm0.9 & 276^{+30}_{-27} & 32\pm3 & 274^{+31}_{-28} \\
CraI & 145 & -0.045\pm0.28\pm0.063 & -0.165\pm0.172\pm0.063 & -0.185 & 149.3\pm1.2 & 112^{+133}_{-77} & -10\pm3 & 112^{+132}_{-79} \\
CraII & 111 & -0.184\pm0.061\pm0.035 & -0.106\pm0.031\pm0.035 & -0.041 & 87.5\pm0.4 & 113^{+24}_{-19} & -84\pm2 & 76^{+34}_{-35} \\
DraI & 79 & -0.012\pm0.013\pm0.035 & -0.158\pm0.015\pm0.035 & 0.131 & -291\pm0.1 & 160^{+19}_{-15} & -89\pm2 & 134^{+22}_{-19} \\
DraII & 24 & 1.242\pm0.276\pm0.057 & 0.845\pm0.285\pm0.057 & -0.591 & -347.6\pm1.8 & 355^{+25}_{-24} & -156\pm8 & 319^{+27}_{-27} \\
EriII & 365 & 0.159\pm0.292\pm0.053 & 0.372\pm0.34\pm0.053 & -0.257 & 75.6\pm2.4 & 617^{+523}_{-393} & -71\pm6 & 612^{+526}_{-401} \\
FnxI & 141 & 0.374\pm0.004\pm0.035 & -0.401\pm0.005\pm0.035 & -0.46 & 55.3\pm0.3 & 138^{+26}_{-25} & -41\pm1 & 132^{+27}_{-27} \\
GruI & 116 & -0.261\pm0.172\pm0.046 & -0.437\pm0.238\pm0.046 & 0.247 & -140.5\pm2.4 & 274^{+102}_{-69} & -196\pm4 & 191^{+127}_{-117} \\
HerI & 129 & -0.297\pm0.118\pm0.035 & -0.329\pm0.094\pm0.035 & 0.136 & 45\pm1.1 & 163^{+31}_{-9} & 152\pm1 & 59^{+61}_{-37} \\
HorI & 83 & 0.891\pm0.088\pm0.058 & -0.55\pm0.08\pm0.058 & 0.294 & 112.8\pm2.6 & 213^{+48}_{-44} & -34\pm3 & 210^{+49}_{-44} \\
HyaII & 148 & -0.416\pm0.519\pm0.061 & 0.134\pm0.422\pm0.061 & -0.427 & 303.1\pm1.4 & 284^{+259}_{-141} & 118\pm6 & 258^{+272}_{-177} \\
HyiI & 26 & 3.733\pm0.038\pm0.035 & -1.605\pm0.036\pm0.035 & 0.264 & 80.4\pm0.6 & 370^{+14}_{-13} & -57\pm2 & 366^{+14}_{-13} \\   
LeoI & 273 & -0.086\pm0.059\pm0.035 & -0.128\pm0.062\pm0.035 & -0.358 & 282.5\pm0.5 & 181^{+44}_{-13} & 167\pm1 & 72^{+79}_{-48} \\
LeoII & 227 & -0.025\pm0.08\pm0.035 & -0.173\pm0.083\pm0.035 & -0.401 & 78\pm0.1 & 77^{+77}_{-44} & 20\pm1 & 74^{+78}_{-49} \\
LeoIV & 155 & -0.59\pm0.531\pm0.059 & -0.449\pm0.358\pm0.059 & -0.237 & 132.3\pm1.4 & 312^{+306}_{-217} & 2\pm8 & 312^{+306}_{-217} \\
LeoV & 174 & -0.097\pm0.557\pm0.057 & -0.628\pm0.302\pm0.057 & 0.047 & 173.3\pm3.1 & 312^{+307}_{-210} & 51\pm7 & 308^{+309}_{-219} \\
PhxI & 419 & 0.079\pm0.099\pm0.04 & -0.049\pm0.12\pm0.04 & -0.162 & -21.2\pm1 & 192^{+158}_{-67} & -117\pm2 & 153^{+176}_{-108} \\
PisII & 182 & -0.108\pm0.645\pm0.061 & -0.586\pm0.498\pm0.061 & 0.053 & -226.5\pm2.7 & 401^{+434}_{-265} & -65\pm8 & 395^{+438}_{-275} \\
RetII & 33 & 2.398\pm0.04\pm0.035 & -1.319\pm0.048\pm0.035 & 0.166 & 62.8\pm0.5 & 248^{+15}_{-14} & -102\pm2 & 226^{+17}_{-16} \\
SgrI & 19 & -2.736\pm0.009\pm0.035 & -1.357\pm0.008\pm0.035 & 0.114 & 140\pm2 & 312^{+21}_{-18} & 142\pm1 & 278^{+23}_{-20} \\
SclI & 85 & 0.084\pm0.006\pm0.035 & -0.133\pm0.006\pm0.035 & 0.157 & 111.4\pm0.1 & 199^{+22}_{-21} & 75\pm1 & 184^{+22}_{-24} \\
Seg1 & 28 & -1.697\pm0.195\pm0.035 & -3.501\pm0.175\pm0.035 & -0.087 & 208.5\pm0.9 & 233^{+28}_{-26} & 116\pm4 & 201^{+31}_{-30} \\
Seg2 & 42 & 1.656\pm0.155\pm0.045 & 0.135\pm0.104\pm0.045 & 0.234 & -39.2\pm2.5 & 224^{+39}_{-34} & 73\pm3 & 212^{+40}_{-35} \\
SxtI & 89 & -0.438\pm0.028\pm0.035 & 0.055\pm0.028\pm0.035 & -0.238 & 224.2\pm0.1 & 242^{+25}_{-22} & 79\pm1 & 229^{+27}_{-23} \\
TriII & 35 & 0.588\pm0.187\pm0.051 & 0.554\pm0.161\pm0.051 & 0.032 & -381.7\pm1.1 & 334^{+20}_{-18} & -256\pm3 & 214^{+29}_{-28} \\
TucII & 54 & 0.91\pm0.059\pm0.035 & -1.159\pm0.074\pm0.035 & -0.374 & -129.1\pm3.5 & 283^{+24}_{-20} & -187\pm2 & 212^{+33}_{-29} \\
TucIII & 21 & -0.025\pm0.034\pm0.035 & -1.661\pm0.035\pm0.035 & -0.401 & -102.3\pm2 & 236^{+5}_{-5} & -228\pm2 & 61^{+12}_{-12} \\
UMaI & 102 & -0.683\pm0.094\pm0.035 & -0.72\pm0.13\pm0.035 & -0.1 & -55.3\pm1.4 & 257^{+62}_{-53} & 10\pm2 & 257^{+62}_{-53} \\
UMaII & 41 & 1.691\pm0.053\pm0.035 & -1.902\pm0.066\pm0.035 & -0.115 & -116.5\pm1.9 & 288^{+21}_{-19} & -59\pm2 & 282^{+21}_{-19} \\
UMiI & 78 & -0.184\pm0.026\pm0.035 & 0.082\pm0.023\pm0.035 & -0.387 & -246.9\pm0.1 & 153^{+17}_{-16} & -71\pm2 & 136^{+19}_{-19} \\
Wil1 & 43 & 0.199\pm0.187\pm0.053 & -1.342\pm0.366\pm0.053 & -0.154 & -12.3\pm2.5 & 120^{+56}_{-44} & 23\pm4 & 118^{+57}_{-47} \\              
                   \noalign{\smallskip}
            \hline
         \end{array}
   $  
   \end{table*}     
            
\section{Velocities and orbital parameters}  

In order to convert the measured proper motion into tangential velocity in an heliocentric reference frame,  
we adopted the  distance modulus listed in Table~\ref{KapSou}, giving priority to distance estimates from studies of variable stars, for those cases where multiple works based on different estimators were available (e.g. mean magnitude of the horizontal branch or tip of the red giant branch). As error in the distance modulus, besides including that from the original sources (see Table~\ref{KapSou}), we added a floor of 0.1\,mag in quadrature, as safeguard against underestimated systematic errors. The line-of-sight and tangential heliocentric velocities are then transformed into velocities in a Cartesian heliocentric (and then Galactocentric) reference system. From these we then calculate orbital poles.

Table~\ref{KapSou2} lists the 3D Galactocentric velocities, $v_{\rm 3D}$, and the Galactocentric radial and tangential velocity components, $v_{\rm rad}$ and $v_{\rm tan}$, respectively; these 
were derived assuming a 8.2$\pm$0.1\,kpc distance of the Sun from 
the Milky Way center and a velocity of the Sun with respect to the Milky Way center of 
(11$\pm$1, 248$\pm$3, 7.3$\pm$0.5) km/s in a reference frame in which the x-axis is positive pointing inwards 
from the Sun to the Galactic center, the y-axis in the 
direction of Galactic rotation and the z-axis towards the North Galactic Cap \citep{Bland_16}. 

 We derive the errors of derived properties from Monte Carlo (MC) simulations. Typically, they are normal "forward" MC simulations, i.e. where the simulated quantities are obtained by random extraction from Gaussian distributions centered on the observed value and with dispersion given by the error in the observed quantities. However, for positive defined quantities such as $V_{\rm tan}$ and $V_{\rm 3D}$ this produces a bias when the errors are close to being as large or larger than the expected values, with the median of the distribution of values from the MC simulations not coinciding with the observed "error-free" value. We take care of this aspect by running "backward" MC simulations. We refer the reader to Appendix~\ref{mc_errors} for the details.

\begin{figure}
   \centering
   \includegraphics[width=0.72\columnwidth,angle=-90]{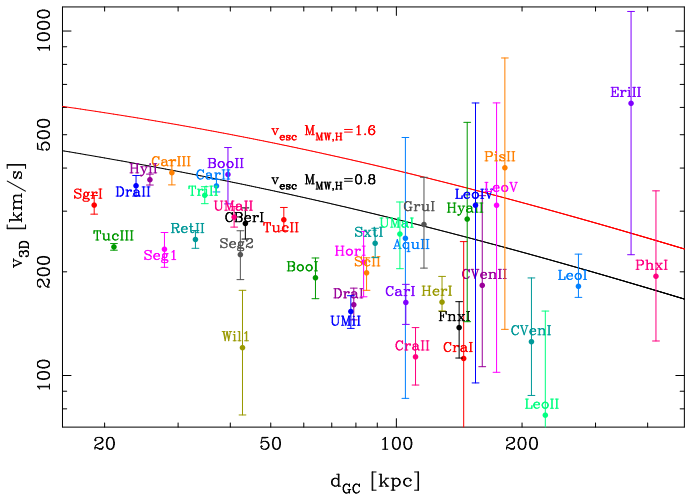}   
      \caption{Total velocity of all galaxies in the sample (points with error bars). 
The curves show the escape velocity for the two potentials used in this paper, i.e. {\it MWPotential14} with a NFW halo of virial mass 0.8$\times 10^{12}$M$_\odot$ (black line) and a more massive 
variant, with virial mass 1.6$\times 10^{12}$M$_\odot$ (red line). 
              }
         \label{vtot}
   \end{figure}

We show the total velocities of all galaxies in Figure~\ref{vtot}. 

We then use the the publicly available code {\it galpy} to integrate orbits in a Galactic potential. 
We adopt the standard {\it MWPotential14} model, which has three components: a spherical bulge, a disc and 
a NFW halo (see Bovy et al. 2015 for details). For the MW dark matter halo mass we explore two cases: a light halo with virial mass 0.8$\times 10^{12}$M$_\odot$ \citep{Bovy_15} and an heavier model, identical to {\it MWPotential14} but for the MW DM halo virial mass, which is 1.6$\times 10^{12}$M$_\odot$ \citep{Fritz_17}. In Table~\ref{KapSou3} we list the apocenter $r_a$, pericenter $r_p$  and eccentricity values derived within these two potentials. The eccentricity is calculated as $ecc=(r_\mathrm{apo}-r_\mathrm{peri})/(r_\mathrm{po}+r_\mathrm{peri})$ within {\it galpy}. 
The distance of the Sun from the Galactic center and the Sun motion are slightly different than what used above, but in previous works it was found that this was causing only minor changes to the orbital 
properties; for our sample, the observational sources of errors are going to be dominant in nearly all cases. 
We use MC realizations of the orbit integrations to estimate the errors on the orbital parameters, see appendix~\ref{mc-back} for the details.

      \begin{table*}
      \caption[]{Orbital properties of objects in our sample. Column 1 lists the object name, cols. 2, 3 and 4  (cols.5, 6, 7) the pericenter, apocenter and eccentricity of the orbit around the Milky Way in the more (less) massive potential of the DM halo with virial mass 1.6$\times10^{12}$ M$_{\odot}$ ($0.8\times10^{12}$ M$_\odot$). 
We note that when the value of $r_a$ minus 1 $\sigma$ is larger than 500 kpc, we only quote " $>500$ kpc" in the table. Our estimates do not include the effect of the gravitational potential of M~31, therefore in general the results for systems with very large apocenters should be taken with a grain of salt  and should be seen as indication that the given object can be on its first infall to the Milky Way.}
         \label{KapSou3}
      $$
         \begin{array}{p{0.15\linewidth}lllllll}
            \hline
          
              \hline
            satellite & \mathrm{peri} (1.6) [kpc] & \mathrm{apo} (1.6) [kpc] & \mathrm{ecc} (1.6) [kpc] & \mathrm{peri} (0.8) [kpc] & \mathrm{apo} (0.8) [kpc] & \mathrm{ecc} (0.8) [kpc] \\
            \hline
            AquII & 95^{+13}_{-79} & 190^{+38581}_{-82} & 0.75^{+0.24}_{-0.45} & 99^{+10}_{-75} & 1212^{+50158}_{-1094} & 0.94^{+0.06}_{-0.52} \\
BooI & 32^{+11}_{-10} & 77^{+9}_{-7} & 0.41^{+0.12}_{-0.08} & 45^{+9}_{-13} & 110^{+51}_{-22} & 0.45^{+0.08}_{-0.03} \\
BooII & 38^{+2}_{-3} & 167^{+624}_{-87} & 0.63^{+0.28}_{-0.27} & 39^{+2}_{-2} & 17206^{+19611}_{-16870} & 0.996^{+0.002}_{-0.2} \\
CVenI & 54^{+87}_{-41} & 254^{+104}_{-23} & 0.66^{+0.24}_{-0.25} & 85^{+87}_{-67} & 325^{+1363}_{-60} & 0.71^{+0.23}_{-0.18} \\
CVenII & 85^{+67}_{-70} & 234^{+8650}_{-47} & 0.71^{+0.27}_{-0.26} & 116^{+39}_{-95} & 464^{+29220}_{-245} & 0.87^{+0.12}_{-0.29} \\
CarI & 60^{+21}_{-16} & 106^{+7}_{-7} & 0.27^{+0.12}_{-0.12} & 103^{+8}_{-23} & 123^{+66}_{-21} & 0.14^{+0.15}_{-0.09} \\
CarII & 26^{+2}_{-2} & 118^{+30}_{-19} & 0.64^{+0.04}_{-0.03} & 29^{+10}_{-3} & >500 & 0.97^{+0.03}_{-0.05} \\
CarIII & 28^{+2}_{-2} & 106^{+63}_{-32} & 0.58^{+0.12}_{-0.11} & 29^{+2}_{-2} & 4445^{+16934}_{-4008} & 0.987^{+0.01}_{-0.1} \\
CBerI & 42^{+3}_{-4} & 63^{+24}_{-14} & 0.2^{+0.13}_{-0.06} & 43^{+3}_{-2} & 183^{+212}_{-72} & 0.62^{+0.18}_{-0.16} \\
CraI & 46^{+97}_{-38} & 153^{+345}_{-11} & 0.63^{+0.31}_{-0.38} & 81^{+65}_{-70} & 159^{+7932}_{-16} & 0.68^{+0.3}_{-0.43} \\
CraII & 18^{+14}_{-10} & 124^{+9}_{-10} & 0.74^{+0.13}_{-0.15} & 27^{+23}_{-15} & 140^{+17}_{-15} & 0.66^{+0.17}_{-0.17} \\
DraI & 28^{+12}_{-7} & 91^{+13}_{-9} & 0.53^{+0.07}_{-0.09} & 42^{+16}_{-11} & 115^{+34}_{-19} & 0.47^{+0.05}_{-0.03} \\
DraII & 19^{+2}_{-2} & 62^{+20}_{-11} & 0.53^{+0.07}_{-0.06} & 20^{+2}_{-1} & 262^{+511}_{-105} & 0.86^{+0.09}_{-0.08} \\
EriII & 356^{+26}_{-45} & >500 & 0.99^{+0.004}_{-0.03} & 357^{+26}_{-37} & >500 & 0.99^{+0.004}_{-0.03} \\
FnxI & 58^{+26}_{-18} & 147^{+9}_{-7} & 0.42^{+0.14}_{-0.13} & 100^{+28}_{-33} & 168^{+55}_{-17} & 0.28^{+0.14}_{-0.05} \\
GruI & 58^{+34}_{-42} & 329^{+9305}_{-130} & 0.81^{+0.17}_{-0.11} & 67^{+30}_{-47} & 8852^{+23824}_{-8446} & 0.986^{+0.009}_{-0.1} \\
HerI & 14^{+23}_{-9} & 187^{+28}_{-21} & 0.85^{+0.1}_{-0.18} & 20^{+32}_{-14} & 284^{+197}_{-46} & 0.87^{+0.09}_{-0.1} \\
HorI & 70^{+19}_{-26} & 94^{+61}_{-15} & 0.21^{+0.18}_{-0.08} & 80^{+12}_{-13} & 206^{+688}_{-106} & 0.44^{+0.38}_{-0.23} \\
HyaII & 116^{+35}_{-89} & 676^{+51725}_{-480} & 0.89^{+0.1}_{-0.36} & 135^{+19}_{-95} & 17518^{+43344}_{-17245} & 0.985^{+0.01}_{-0.28} \\
HyiI & 25^{+2}_{-1} & 73^{+18}_{-12} & 0.49^{+0.06}_{-0.05} & 25^{+2}_{-1} & 451^{+799}_{-185} & 0.89^{+0.06}_{-0.06} \\
LeoI & 45^{+80}_{-34} & 590^{+584}_{-90} & 0.87^{+0.09}_{-0.09} & 63^{+221}_{-47} & >500 & 0.96^{+0.02}_{-0.03} \\
LeoII & 41^{+125}_{-30} & 238^{+115}_{-22} & 0.73^{+0.2}_{-0.44} & 67^{+154}_{-52} & 248^{+613}_{-26} & 0.67^{+0.26}_{-0.39} \\
LeoIV & 150^{+10}_{-112} & 1794^{+64581}_{-1637} & 0.95^{+0.05}_{-0.62} & 153^{+8}_{-87} & 26071^{+46619}_{-25908} & 0.989^{+0.007}_{-0.56} \\
LeoV & 165^{+14}_{-126} & 4079^{+63439}_{-3891} & 0.96^{+0.04}_{-0.58} & 168^{+12}_{-104} & 27704^{+45671}_{-27495} & 0.988^{+0.007}_{-0.46} \\
PhxI & 263^{+126}_{-219} & >500 & 0.91^{+0.07}_{-0.23} & 302^{+91}_{-236} & >500 & 0.96^{+0.02}_{-0.11} \\
PisII & 171^{+24}_{-102} & 31214^{+67612}_{-30994} & 0.992^{+0.005}_{-0.32} & 173^{+24}_{-60} & 41983^{+61249}_{-41684} & 0.992^{+0.005}_{-0.32} \\
RetII & 23^{+4}_{-3} & 43^{+6}_{-4} & 0.31^{+0.02}_{-0.02} & 27^{+3}_{-3} & 76^{+27}_{-14} & 0.47^{+0.07}_{-0.04} \\
SgrI & 15^{+2}_{-2} & 36^{+9}_{-6} & 0.42^{+0.03}_{-0.02} & 16^{+2}_{-2} & 79^{+45}_{-23} & 0.67^{+0.08}_{-0.06} \\
SclI & 51^{+15}_{-10} & 100^{+17}_{-9} & 0.32^{+0.07}_{-0.04} & 69^{+10}_{-9} & 169^{+105}_{-40} & 0.42^{+0.14}_{-0.05} \\
Seg1 & 16^{+4}_{-3} & 36^{+6}_{-4} & 0.39^{+0.06}_{-0.04} & 20^{+4}_{-3} & 54^{+22}_{-12} & 0.48^{+0.07}_{-0.04} \\
Seg2 & 29^{+8}_{-8} & 49^{+14}_{-6} & 0.27^{+0.09}_{-0.04} & 37^{+5}_{-6} & 80^{+75}_{-24} & 0.38^{+0.18}_{-0.08} \\
SxtI & 71^{+11}_{-12} & 131^{+50}_{-24} & 0.3^{+0.07}_{-0.02} & 79^{+9}_{-8} & 419^{+810}_{-181} & 0.68^{+0.19}_{-0.15} \\
TriII & 16^{+3}_{-3} & 92^{+23}_{-13} & 0.71^{+0.02}_{-0.02} & 19^{+3}_{-3} & 415^{+824}_{-165} & 0.91^{+0.05}_{-0.03} \\
TucII & 29^{+8}_{-7} & 107^{+36}_{-23} & 0.58^{+0.03}_{-0.02} & 34^{+7}_{-7} & 345^{+562}_{-144} & 0.82^{+0.09}_{-0.06} \\
TucIII & 2^{+1}_{-1} & 33^{+3}_{-2} & 0.86^{+0.03}_{-0.03} & 3^{+1}_{-1} & 47^{+5}_{-4} & 0.89^{+0.02}_{-0.03} \\
UMaI & 100^{+8}_{-13} & 175^{+306}_{-72} & 0.31^{+0.34}_{-0.22} & 101^{+7}_{-7} & 1174^{+19579}_{-940} & 0.84^{+0.15}_{-0.43} \\
UMaII & 38^{+3}_{-4} & 65^{+22}_{-12} & 0.26^{+0.1}_{-0.05} & 39^{+3}_{-3} & 196^{+211}_{-68} & 0.67^{+0.15}_{-0.11} \\
UMiI & 29^{+8}_{-6} & 85^{+7}_{-7} & 0.49^{+0.07}_{-0.08} & 44^{+12}_{-10} & 101^{+17}_{-11} & 0.39^{+0.06}_{-0.04} \\
Wil1 & 12^{+13}_{-6} & 43^{+9}_{-8} & 0.54^{+0.21}_{-0.24} & 18^{+23}_{-10} & 44^{+13}_{-8} & 0.42^{+0.27}_{-0.24} \\

                   \noalign{\smallskip}
            \hline
         \end{array}
     $$                     
   \end{table*}

Figure~\ref{peri-apo} shows the orbital properties of the objects in the sample, quantified as the 
pericenter versus apocenter (for the heavy and light 
NFW halo in the top and bottom panel, respectively).  We only plot objects with an error in 3D velocity of less than 100 km/s: this is less than half the escape speed expected for the low-mass MW-halo at the largest distances probed (see Fig.~\ref{vtot}) and avoids biases towards large values of apocenter for galaxies with larger errors.

In Fig.~\ref{VPOS} we plot an all-sky view of the orbital poles of the objects in the sample, concentrating on those 
within 200\,kpc, and comparing their location on this plane with the
vast polar structure (VPOS) of satellites \citep{Pawlowski_12}.

\section{Results and discussion}

\subsection{Apocenter Distances and Orbital Energies}

In Figure~\ref{vtot} we show how the newly measured 3D velocities of the satellites as a function of
radius compare to the local escape velocity in the two MW potentials.    All the galaxies analyzed are compatible with being bound to the MW, in both the "light" and "heavy" MW potentials. This includes also very distant systems like Phoenix and Eridanus~II, although the large error-bars could place them away from the locus of bound galaxies.  
The proper motion errors for both are still too large to change the result based on LOS velocities, that Eri~II is likely bound \citep{Li_17} and that Phoenix~I is likely on first infall \citep{Kacharov_17}. 

\begin{figure}
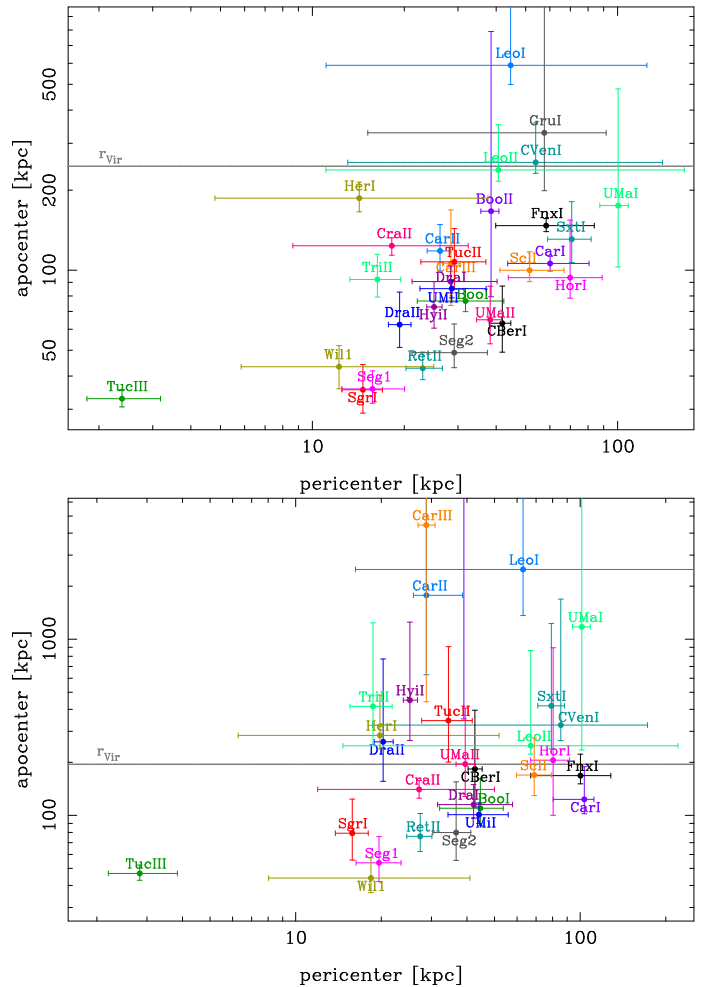

   \centering
   \includegraphics[width=0.72\columnwidth,angle=-90]{peri_apo3_rev1.eps}
   \includegraphics[width=0.72\columnwidth,angle=-90]{peri_apo4_rev1.eps}   
      \caption{Pericenter versus apocenter properties of the galaxies in the sample with 3D velocity errors $<$100 km/s, 
as indicated by the labels. 
The top panel plot shows the results for a MW DM halo of virial mass equal to $1.6\times10^{12}$ M$_\odot$, 
and the lower panel for $0.8\times10^{12}$ M$_\odot$. We also show typical virial radii in both plots. Gru\,I is omitted in the lower plot because their very large apocenter.
              }
         \label{peri-apo}
   \end{figure}

Figure~\ref{peri-apo} allows us to see the number of galaxies that have orbits that
will take them beyond the virial radius of the MW DM halo, i.e. those that have
apocenters above the solid lines.  Such satellites can be considered alongside
the backsplash population of satellites that exist beyond the virial
radius at z$=$0 in simulations, but have been within the virial radius at earlier times
\citep{balogh00,gill05,teyssier12,Simpson_18}.  Considering only the objects with error in $v_{\rm 3D} < 100\,$km/s (30 systems), Leo~I is presently located beyond the virial radius in both MW potentials, while CVen~I and Leo~II for the light MW halo. When considering the values within the 1$\sigma$ range, only Leo~I is found to be certainly "backsplashing" in the heavy halo model, with other possible candidates being CVen~I, Leo~II, UMa~I, Gru~I and Boo~II; the latter two might fall in the category of systems infalling for the first time, given their negative Galactocentric radial velocities.

In the low mass halo, 13 objects have apocenters that would still exceed the virial radius even when subtracting the 1\,$\sigma$ errors and are therefore “backsplashing” satellites
in that their orbits will take them beyond the virial radius (5 of these have negative $v_{\rm rad}$); several others have most likely apocenters smaller than the virial radius but could potentially be scattered beyond it due to the measurement uncertainty. This would imply that a large fraction of the observed dwarfs could be backsplashing, whilst for a high mass
halo the fraction is very much reduced.

The definition of whether a satellite is bound
or not in an evolving cosmological context is complex, and the comparison to the curves in Fig.~1  should
therefore be interpreted primarily in a relative sense. \citet{Boylan-Kolchin_13} addressed this topic in detail for the case of Leo~I. 
That analysis was based
on the proper motion of Leo~I measured by HST, which is in good agreement with our \textit{Gaia} DR2 value.
They showed that the low-mass MW is actually ruled out at 95\% confidence, since in such halos it is
vanishingly rare in cosmological simulations to find subhalos at 273\,kpc moving as fast as Leo~I. By contrast,
they showed that the high-mass Milky Way is the statistically preferred value.  
Given this result, our Figure~\ref{vtot} and Figure~\ref{peri-apo} show that
several other satellites (e.g. Car\,II, Car\,III, Gru\,I) have a combination of distance and velocity that will put them at odds with a low-mass MW.

   \begin{figure*}
   \centering
   \includegraphics[width=0.48\textwidth,angle=0]{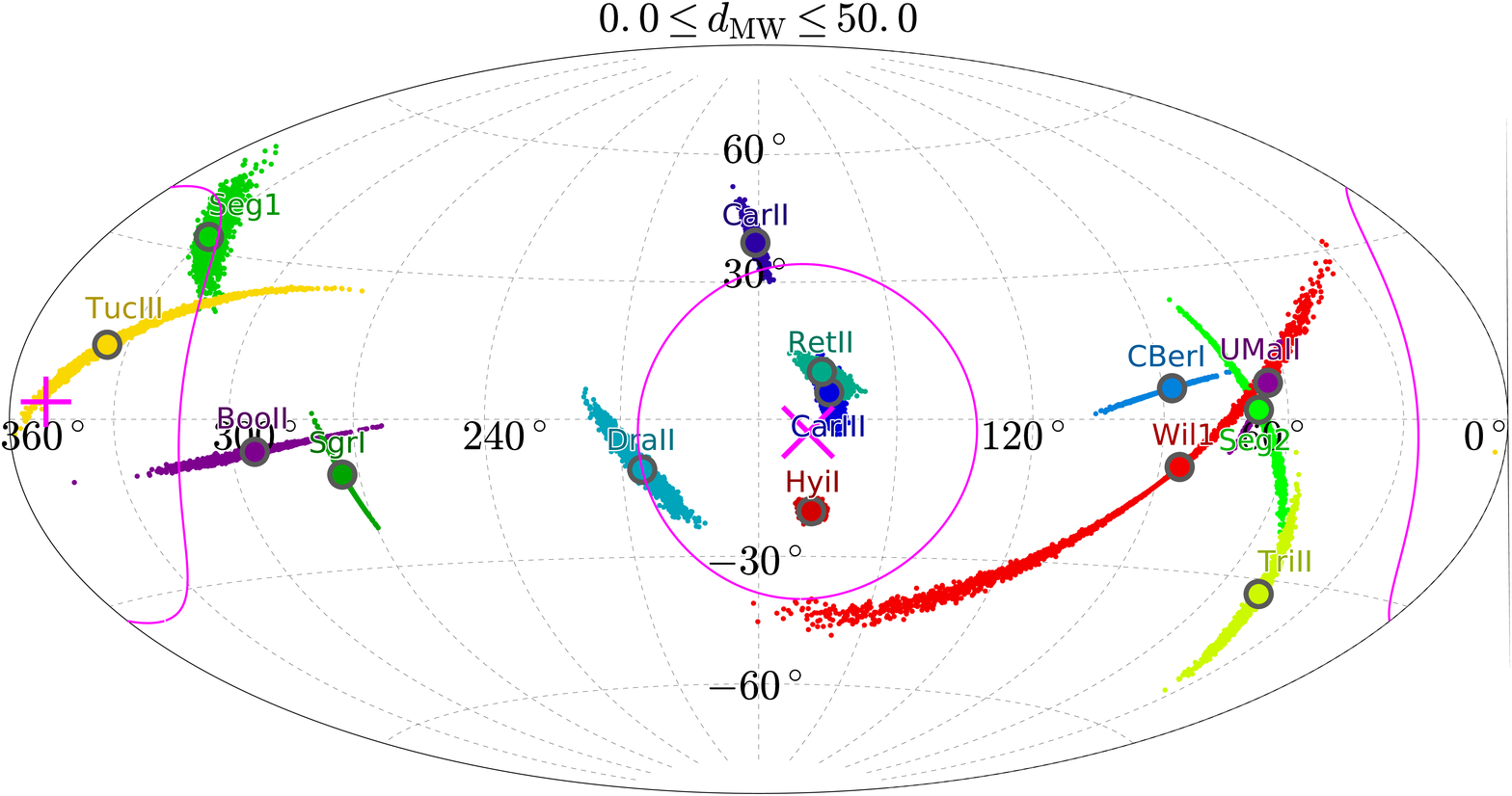}
   \includegraphics[width=0.48\textwidth,angle=0]{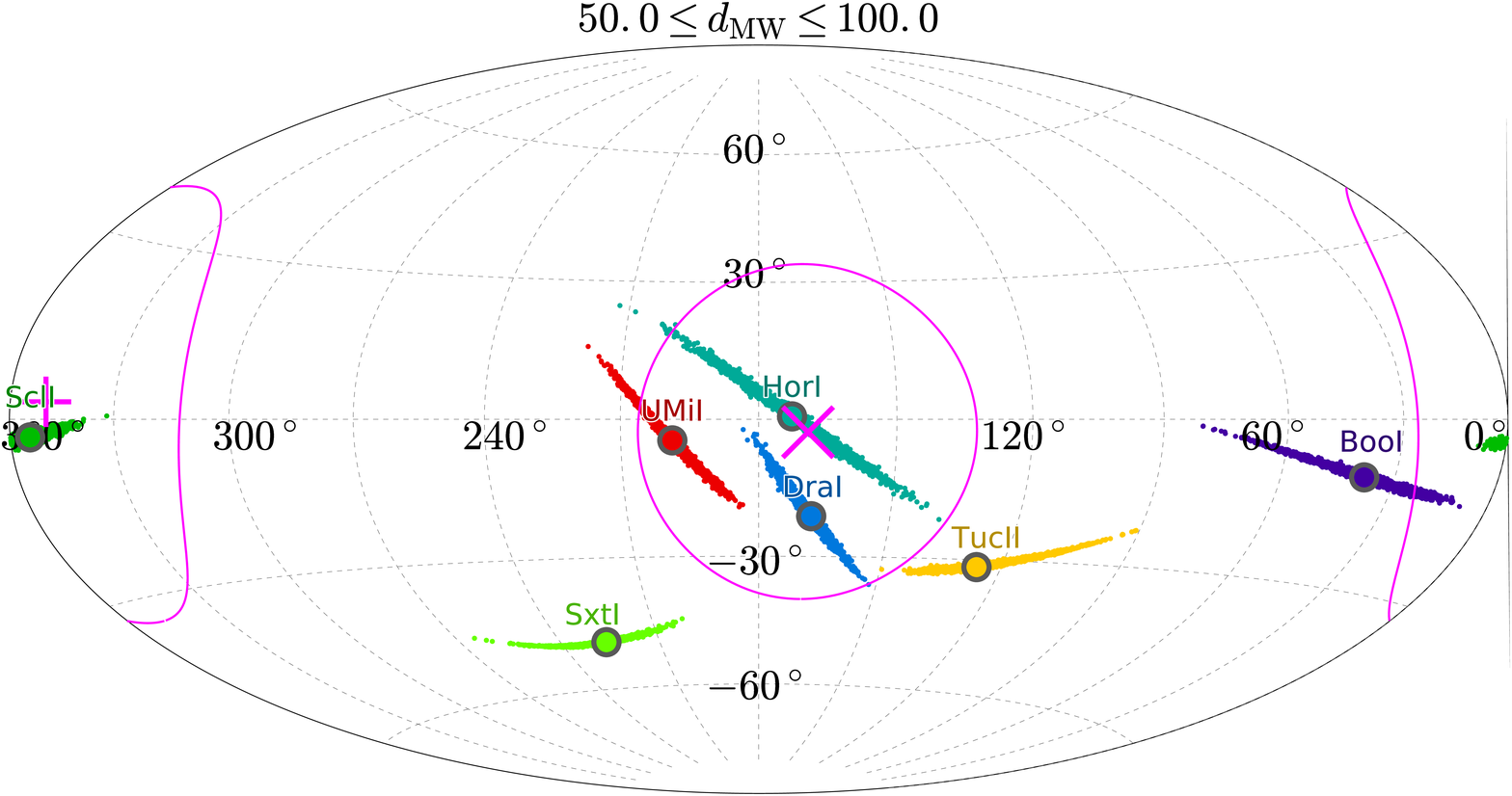}
   \includegraphics[width=0.48\textwidth,angle=0]{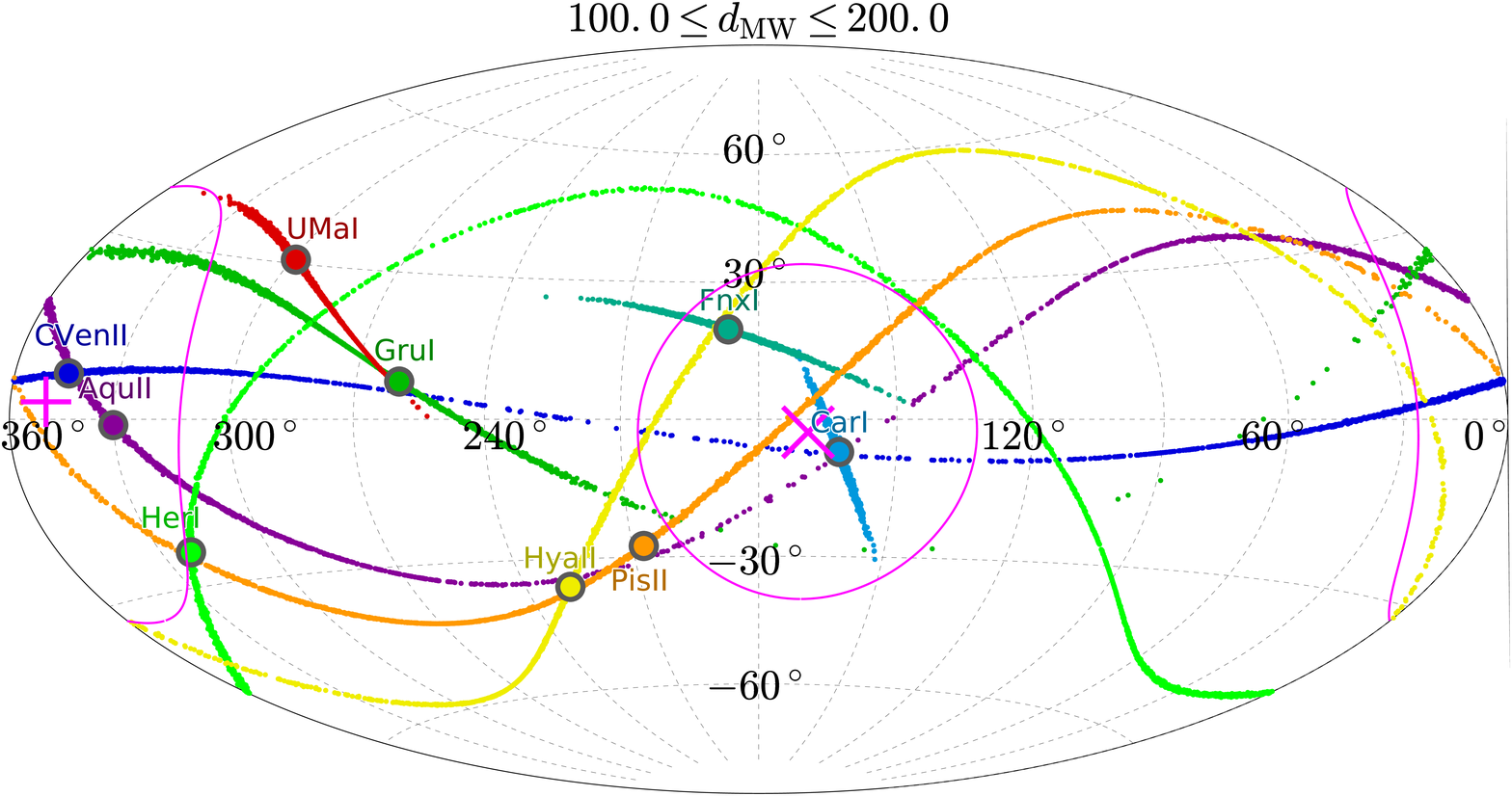}
   \includegraphics[width=0.48\textwidth,angle=0]{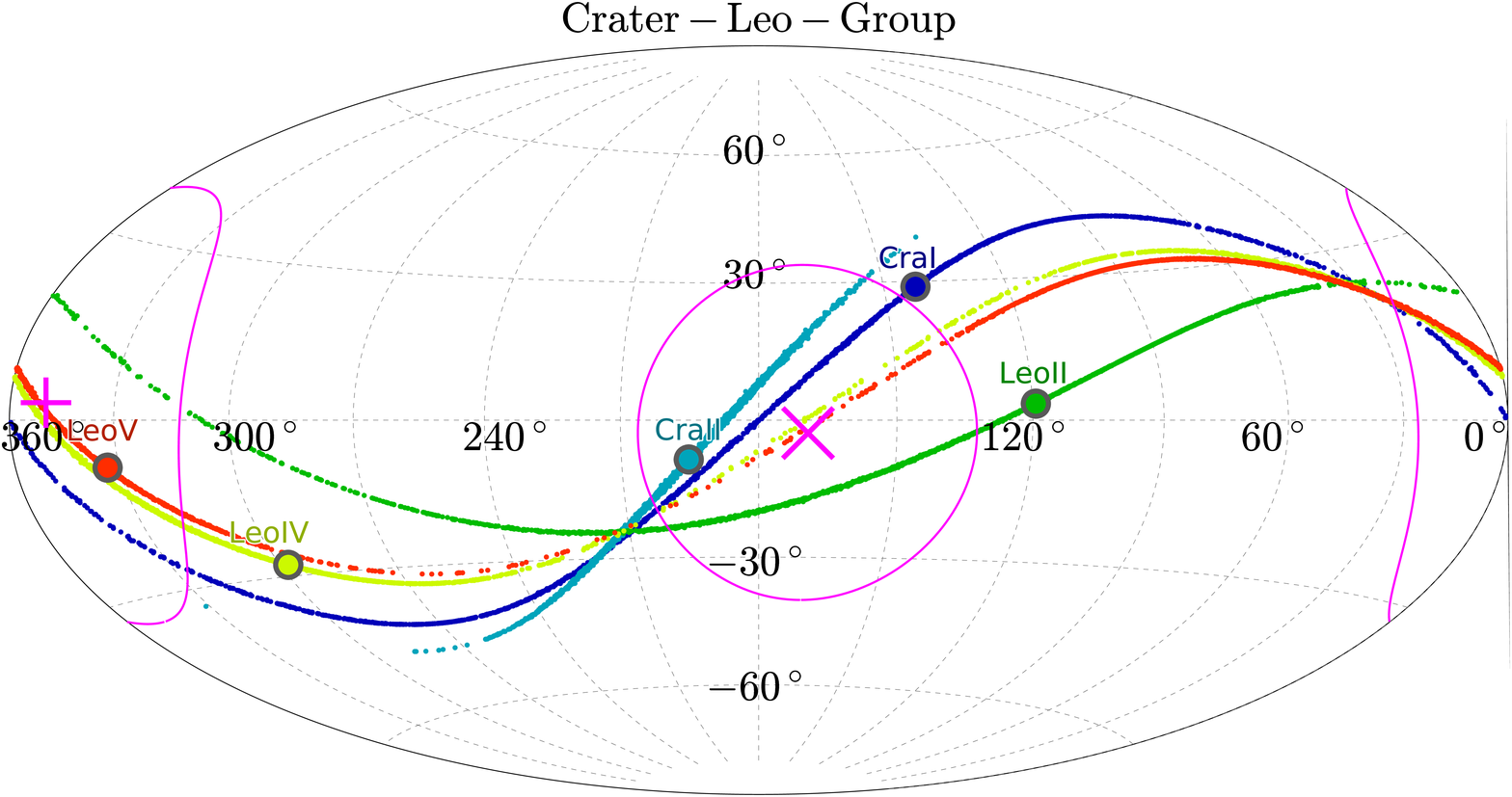}   
      \caption{All-sky view of orbital poles for the objects in the sample; the circles indicate the median of the 
2000 Monte Carlo simulations while the small points around each object plot the orbital poles from the individual 
simulations. The magenta circles contain 10 per cent of the sky around the assumed VPOS pole, which is given as a ``X'' for the “co-orbiting” direction (orbital sense as most classical satellites, including the LMC and SMC), and a ``$+$'' for the opposite normal direction (“counter-orbiting”). Top panels: objects with Galactocentric distances between 0-50\,kpc (left) and 50-100\,kpc (right); bottom panels: objects with Galactocentric distances between 100-200\,kpc (left) and in the putative Crater-Leo group.
              }
         \label{VPOS}
   \end{figure*}

   \begin{figure}
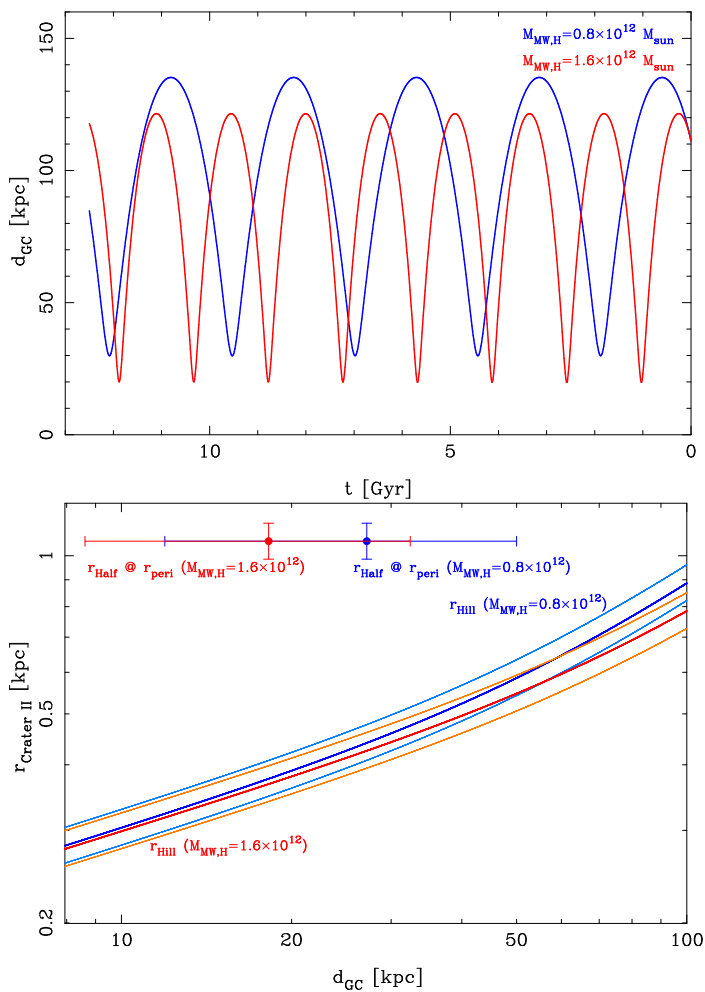

   \centering
   \includegraphics[width=0.72\columnwidth,angle=-90]{crater2_r_t.eps}
   \includegraphics[width=0.72\columnwidth,angle=-90]{peri_hill_rev1.eps}
      \caption{Properties of Crater\,II. Top: Orbital integration for Cra\,II in the {\it MWpotential14} for the standard 
light halo (blue line) and the heavy DM halo (red line) plotted as the 
distance from the MW center as a function of time. Bottom: Hill radius of a galaxy of the mass of Cra\,II \citep{Caldwell_17} in two different MW potentials compared to the size and apocenter of Cra\,II.  
}
         \label{crater2}
   \end{figure}

\subsection{Pericenter Distances and Tidal Influences}
From Figure~\ref{peri-apo} it can be noticed that the measured proper motion of Tuc~III leads to a very internal orbit in both MW potentials, with the object reaching well within 10\,kpc and being confined within 50\,kpc from the MW center. Tuc~III is therefore highly likely to have been subject to strong tidal disturbances, as confirmed by the presence of a stream around it and predicted 
by \citet{Erkal_18}. Nonetheless, the proper motion we measure is off from the predicted ones in  \citet{Erkal_18}, independently on their adopted LMC mass; this could be indicative  of a difference between the MW potential assumed in that work and the actual MW potential. We note that Tuc~III is one of the few satellites in our sample whose nature as a galaxy is not confirmed according to the spread in the distribution of l.o.s. velocities and metallicities \citep{Simon_17,Li_18b}; perhaps its possible nature as a stellar cluster could explain why its pericenter is much smaller pericenter than all of the other systems.

The following galaxies have a likely pericenter smaller than 20 kpc in at least one of the MW potentials: Wil\,1, Seg\,1, Sgr\,I, Cra\,II, Dra\,II, Tri~\,I and Her\,I. This is clearly a dangerous region, as empirically highlighted by the obvious state of tidal disruption afflicting Sgr~I \citep{Majewski_03}. The other galaxies broadly consist of two groups. The first one contains relatively bright galaxies (Cra~II, Her~I), which do also show signs of tidal disruptions (see below). The second group are very faint satellites which, due to selection biases, can only be detected if currently relatively close to us, and thus have a higher probability to have also a smaller pericenter (Wil\,1, Seg\,1, Dra\,II, Tri\,II). Apart from Dra~II \citep{Longeard_18}, for which efficient cleaning of foreground contamination was done through the use of narrow-band filters, no signs of tidal disturbance have been detected yet in the other three systems. This could be due to the intrinsic difficulty in unveiling such signs in extremely faint systems in the presence of contamination, or could also be a consequence of their relative compactness (with half-light radii of at most 21\,pc \citep{Carlin_17}).  We note that, within the more massive MW potential, the galaxies with observed tidal features have closer pericenter passages with respect to the case of the less massive MW, another argument in favor of a more massive halo. The inhomogeneous nature of galaxies found to have small pericenter distances might also be  partially caused by large uncertainties in the derived quantities.

As mentioned in the Introduction, elongations and/or isophote twists suggestive of tidal stretching of the stellar 
component have been observed in Carina\,I, Hercules and Bo\"{o}tes\,I.
For Her~I a pericenter as small as 5-10\,kpc is within the 1\,$\sigma$ bounds for both potentials, thus it supports the nature of the observed elongations being of tidal origin. Our proper motion is within 0.7/1 $\sigma$ of the proper motion prediction of \citet{Kuepper_17}, who explain the structure of Hercules on an orbit with a pericenter distance of about 5\,kpc.

On the other hand Carina~I is not expected to have come closer 
than $\sim$50\,kpc (within 1$\sigma$) for the heavy halo, while in the case of the light halo its orbit would be 
rather external. The observational findings of signs of likely tidal disturbance in Carina can be 
considered as robust as they have been detected in multiple 
studies in the literature adopting different methodologies \citep[see e.g.][and references in]{Munoz_06, Battaglia_12, McMonigal_14}; 
their presence might then suggest a 
preference for a heavy MW DM halo or that dwarf galaxies might experience strong tidal disturbance even on 
rather external orbits. We deem this latter hypothesis unlikely 
\citep[see results from N-body simulations, e.g.][for those simulated objects on similar orbits]{Munoz_08, Penarrubia_08}, 
but this is certainly an important aspect to be verified, since several of the 
MW dwarf galaxies are on more internal orbits than Carina (e.g. Draco, Ursa Minor to mention a few) 
and could see their internal kinematic properties 
potentially affected.  We note though that the hypothesis put forward by \citet{Fabrizio_16} of Carina being the result of an initially disky and rotating dwarf galaxy heavily tidally stirred by the Milky Way would require it to be on an exactly prograde and  tighter (125\,kpc/25\,kpc apocenter/pericenter) orbit than measured here even with the heavy MW potential, and it does not therefore seems likely.  Alternatives could contemplate the possibility that the properties of 
Carina might have been affected by the Magellanic system: e.g. the recent infall of the Magellanic system 
might have modified Carina's current orbital properties, 
making them therefore not representative of its past orbital history; however Carina shares with the LMC 
a very similar orientation of the time-average orbit \citep{Helmi_18}, while torques from the LMC would be expected 
to be maximal on systems on a perpendicular orbital plane; we speculate that Carina could have experienced a close 
encounter with the Magellanic Clouds and that perhaps these latter systems are 
responsible for inducing tidal disturbances.  While the perturbative influence of the LMC \citep{Gomez_15} has not been included in the presented
orbit calculations, it will be included in follow-up studies to assess the
impact on the orbits of the classical and UFD satellites of the MW
(Patel. et al. in prep.).

Depending on the potential adopted, Bo\"{o}tes~I case is intermediate between Car\,I and Her\,I, with a pericenter $\gtrsim$20\,kpc  but for which smaller values cannot be excluded \citep[see also][]{Helmi_18,Simon_18}. Thus, the observed elongated and structured spatial distribution of its stars could still be compatible with tidal features caused by the MW.

Crater\,II is a remarkable galaxy, because it is larger than Fornax but only as luminous as CVen\,I \citep{Torrealba_16b} and has a very cold internal kinematics when compared to galaxies of similar size \citep{Caldwell_17}. Its peculiar properties have raised interest in the community. Its low l.o.s. velocity dispersion appears compatible with the predictions given by MOND based only on the characteristics of Cra\,II stellar component \citep{McGaugh_16}, at least under the assumption of dynamical equilibrium. On the other hand, its internal kinematical and structural properties could also be consistent  with Crater~II having been embedded in a typical dark matter halo  expected in $\Lambda$CDM for dwarf galaxies, which has undergone heavy tidal stripping \citep[see e.g.][]{Fattahi_17, Sanders_18}. 

In Figure~\ref{crater2} (top) we plot Cra\,II distance from the Galactic center as a function of time: in both MW potentials, the best Cra\,II orbit is radial, with an eccentricity of about 0.7, and can reach as close as 2\,~kpc from the MW center.  That implies that the Hill radius $r_\mathrm{Hill}=(m_\mathrm{CraterII}/(3*m_\mathrm{MW}))^{1/3}*r_\mathrm{peri, CraterII}$ (where $m_{\rm MW}$ is the MW mass within the pericenter distance of Crater~II, $r_{\rm peri, CraterII}$)
is far smaller than Cra\,II half-light radius 
(see Figure~\ref{crater2}, bottom panel), therefore confirming the prediction of  \citet{Fattahi_17} and \citet{Sanders_18} that it is in process of tidal disruption. 
Our proper motion is in  good (0.3/1.3$\sigma$) agreement with the 
models by \citet{Sanders_18}.

As stated above, several other systems do share internal orbits such as Crater~II and therefore the potential impact of 
tidal disturbances needs to be understood to properly interpret these systems' structural and internal kinematic properties.

\subsection{The Missing Satellite Problem}

\begin{figure}
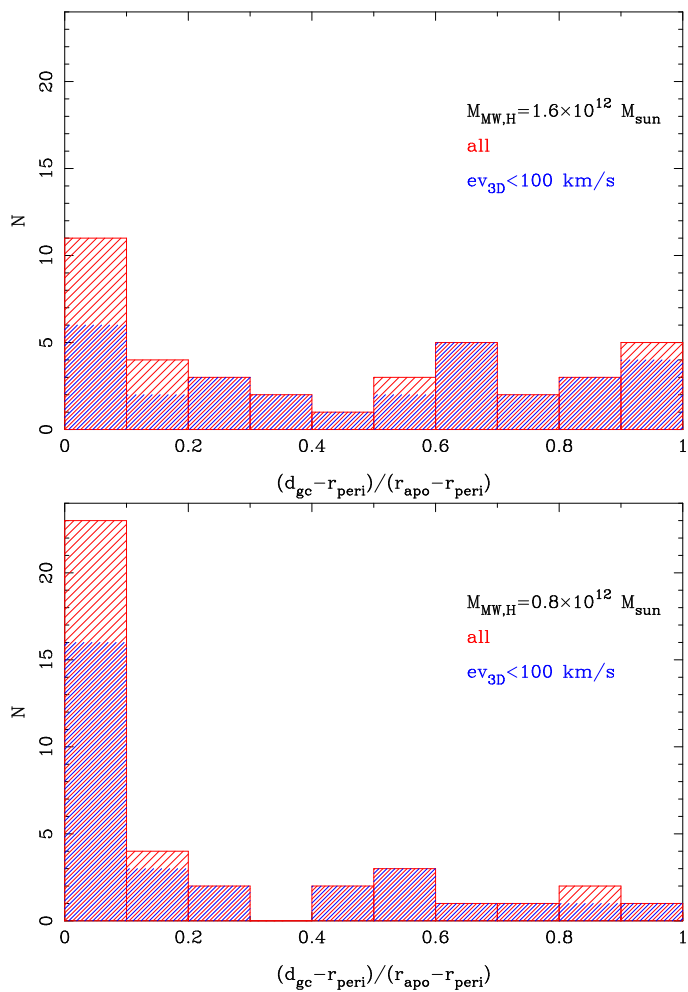

   \centering
   \includegraphics[width=0.72\columnwidth,angle=-90]{his_h16b.eps}
   \includegraphics[width=0.72\columnwidth,angle=-90]{his_h08b.eps}
      \caption{ Histograms of the ratio $f = (d_{\rm
GC} - r_{\rm peri}) / (r_{\rm apo} - r_{\rm peri})$, for both the
high-mass and low-mass MW models (upper and lower panels,
respectively).  A galaxy on the left of each histogram is near
pericenter, while a galaxy on the right of each histogram is near
apocenter.}
         \label{mis-sat}
   \end{figure}

Figure~\ref{mis-sat} shows histograms of the ratio $f = (d_\mathrm{
GC} - r_\mathrm{peri}) / (r_\mathrm{apo} - r_\mathrm{peri})$:   a galaxy with $f$ close to zero   is near
pericenter, while a value close to 1 indicates that the galaxy is near apocenter.  Both histograms have a (weak) peak at small value of $f$.  This is
more pronounced for the low-mass MW model, which based on the
arguments in the preceding sections, we consider less plausible. We also show the variant where only galaxies with $v_\mathrm{3D}<100$ are plotted, since we expect their orbital properties not to suffer from significant biases. When galaxies with large errors are excluded, the histogram for the higher mass halo is close to flat.
However, basic
dynamics dictates that, within their orbits, galaxies spend most of
their time near apocenter, where the velocity is lower. 
If the orbits would be circular the argument would not be valid, but the median eccentricity is 0.53 for the high mass MW halo, when the galaxies with larger errors are excluded.
 
Also, the
number of galaxies at small pericenters ($\lesssim 20$ kpc) is reduced
through tidal destruction.  So if we had a {\it complete} sample of MW
dwarfs, then the histograms would have to be increasing towards high
$f$. By contrast, even for the high-mass MW, the observed histogram is
flat at best. 
One could argue that the peak towards lower values of $f$ could be caused by a group of former satellites of a larger galaxy (e.g. the LMC) having infallen; however, the number of objects that we find as possibly having been associated to each other  is not sufficient to explain the feature (see Section~\ref{Group-infall}).
The corollary is that there must be a population of
(ultra-faint) dwarf galaxies that are currently at apocenter,
especially beyond $\sim 100$ kpc, that have yet to be discovered. This
is relevant to the so-called "missing satellite” problem (see review
in \citet{Kravtsov_09}), as it affects
the comparison of observed dwarf galaxy counts to subhalos found in
hierarchical galaxy formation scenarios. Implications of this
statement will be discussed in Patel et al. (in prep.).

\subsection{Orbital Poles and Planar Alignments}

Figure~\ref{VPOS} shows the distribution of orbital poles for three distance bins (0 to 50, 50 to 100, and 100 to 200 kpc), as well as for the proposed members of the Crater-Leo Group (not included in the 100 to 200 kpc plot). The uncertainty in the direction of orbital poles is illustrated with point clouds based on the 2000 Monte Carlo simulations of the measurement uncertainties. The magenta cross and plus sign give the assumed normal direction to the VPOS, as used to predict the proper motions of satellites in \citet{Pawlowski_13}. The magenta circles of opening angle $\theta_{\mathrm{inVPOS}} = 36.87^\circ$ denote areas of 10 per cent of the sphere around these directions. We consider orbital poles that lie within this region to orbit along the VPOS.

\begin{figure}
   \centering
   \includegraphics[width=0.51\textwidth,angle=0]{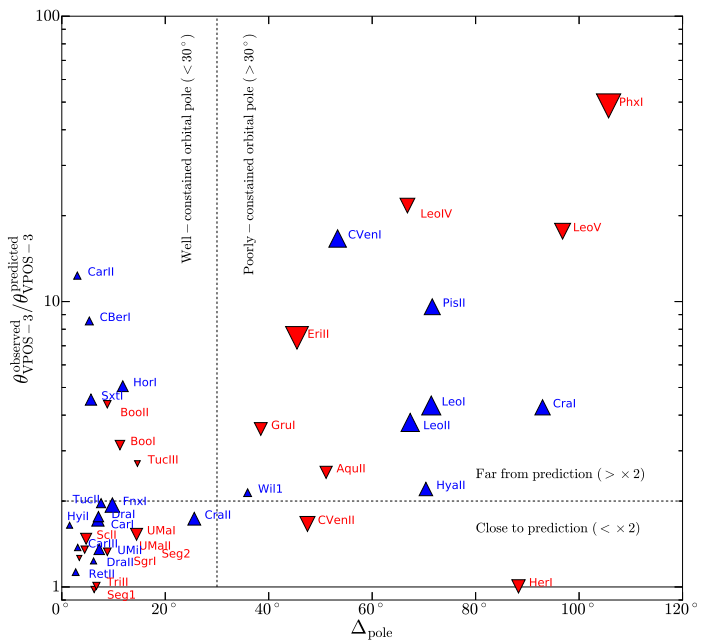}
      \caption{
Alignment with the VPOS, measured as the ratio between the measured ($\theta_{\mathrm{obs}}$) and the predicted ($\theta_{\mathrm{pred}}$) angular offset of an orbital pole from the VPOS, plotted against the uncertainty ($\Delta_{\mathrm{pole}}$) in the direction of the orbital pole. Blue, upward triangles are co-orbiting relative to the majority of classical satellites including the LMC, red downward triangles are counter-orbiting. Symbol size increases with Galactocentric distance.
      }
         \label{VPOS3}
   \end{figure}

Based on the assumption that the satellite galaxies orbit along the VPOS, \citet{Pawlowski_13,Pawlowski_15} predicted orbital poles for these systems. We can now test how well these predictions are met. The predicted orbital pole direction is the direction along the great circle perpendicular to the satellite (as seen from the center of the Milky Way) which minimizes the angle to the assumed VPOS normal, which points to Galactic coordinates $(l,b) = (169.3^\circ, -2.8^\circ)$. The corresponding minimum angle between the VPOS normal and the pole of a satellite galaxy is $\theta^{\mathrm{pred}}$. The angle between the orbital pole based on our measured proper motions and the VPOS normal is $\theta^{\mathrm{obs}}$. The ratio between these gives a measure of how well an observed orbital pole agrees with its predicted direction. Figure~\ref{VPOS3} plots this ratio against a measure of the uncertainty in observed orbital pole direction, $\Delta_{\mathrm{pole}}$. The latter is defined as the angle from the most-likely measured orbital pole of a satellite which contains 68 per cent of the orbital poles sampled from its measurement uncertainties. The ratio of observed to measured angle from the VPOS is typically large for objects with poorly constrained orbital pole directions ($\Delta_{\mathrm{pole}} > 30^\circ$), but for those with smaller uncertainties the majority of objects (17 out of 24) agree to within a factor of two (dashed magenta line) with the predicted angle.

Table~\ref{VPOS2} provides an overview of the predicted ($\theta^{\mathrm{pred}}$) and the measured ($\theta^{\mathrm{obs}}$) alignments of satellite orbital poles. The observed angle is given a negative sign if the measured orbital pole is counter-orbiting relative to the orbital direction of the majority of the classical satellite galaxies associated with the VPOS. For each satellite, we calculate the fraction $p_{\mathrm(inVPOS)}$\ of Monte Carlo sample orbital pole directions which falls to within $\theta_{\mathrm{inVPOS}}$\ of the VPOS normal. For an orbital pole that is misaligned with the VPOS, this indicates whether there nevertheless is a chance that the pole might be aligned.
Also given in the table is a measure of how strong the constraints of each satellite's pole are on its alignment with the VPOS direction. For the latter, we assume that the satellite's intrinsic orbital pole aligns perfectly with its predicted direction. We then vary its orbital pole direction by sampling 2000 times from the measured uncertainties in orbital pole direction of this satellite. Since a satellite's orbital pole can not be better aligned than the predicted direction, any offset from this direction results in an increase of the angle with the VPOS normal. For each realization, the angle to the VPOS normal is calculated. If it is larger than $\theta_{\mathrm{inVPOS}}$, we count this realization as not aligned with the VPOS, even though we know that its intrinsic pole is perfectly aligned. The fraction of realizations counted this way gives an estimate of the probability $p_{\mathrm{>VPOS}}$\ to falsely find this satellite's orbital pole to be misaligned with the VPOS. It is compiled in column 5 of the table. We also count how often the angle to the VPOS exceeds $\theta^{\mathrm{obs}} [^\circ]$, the angle between the most-likely measured orbital pole and the VPOS normal. This gives an estimate of the probability $p_{\mathrm{>obs}}$\ to measure an intrinsically well aligned orbital pole as far away from the VPOS normal as observed. Since this method assumes intrinsically perfect alignments, the resulting probabilities should be seen as lower limits.

Overall, for 12 of the satellites the chance $p_{\mathrm{inVPOS}}$\ to align with the VPOS to better than $\theta_{\mathrm{inVPOS}}$\ given our proper motion uncertainties is lower than 5 per cent. For the remaining 27 satellites, an alignment of their orbital pole with the VPOS is either found, or can not be rejected with high confidence.
Six out of the 39 objects can not align with the VPOS because their predicted angle $\theta^{\mathrm{pred}}$\ already exceeds $\theta_{\mathrm{inVPOS}}$: their spatial positions alone already place them outside of the VPOS plane orientation. A well known example is Sagittarius, which has an orbit almost perpendicular to the VPOS. The other satellites in this category are Hercules\,I, Segue\,2, Triangulum\,II, Ursa Major\,II, and Wilman\,1. Five of these counter-orbit relative to the VPOS, for a counter-orbiting fraction of $f_\mathrm{counter} = \frac{5}{6} = 0.83$. Of the remaining 33 satellites for which an alignment is feasible, 17 have median orbital poles which align to within $\theta_{\mathrm{inVPOS}}$\ with the VPOS normal. The majority of these are well constrained to align with the VPOS, because most of their Monte Carlo sampled orbital poles also shows an alignment ($p_{\mathrm{inVPOS}} > 0.5$). The three least certain alignments are CVen\,I, Leo\,V, and Dra\,II, for which only 38, 40, and 48 per cent of realizations fall to within $\theta_{\mathrm{inVPOS}}$\ of the VPOS normal, respectively. Of these 17 satellites with aligned orbital poles, only one has $p_{\mathrm{>VPOS}} \geq 0.5$\ (Leo\,V), five have $0.4 \geq p_{\mathrm{>VPOS}} \geq 0.17$, and the others have $p_{\mathrm{>VPOS}} < 0.1$. This indicates that given the respective measurement uncertainties, the measured orbital poles are expected to be found within $\theta_{\mathrm{inVPOS}}$\ of the VPOS if the satellites orbit along the VPOS. It is also interesting to note that of these 17 satellites, six counter-orbit. These are Aqu\,II, CVen\,II, Leo\,V, Seg\,1, Tuc\,III and Scu\,I, for which this was known previously). The remaining 11 satellites co-orbit, as do the LMC and SMC, so the counter-orbiting fraction is $f_\mathrm{counter} = \frac{6}{17 + 2} = 0.32$.

The remaining 16 satellites have median orbital poles which do not align with the VPOS. Of these, nine have orbital pole directions that are too weakly constrained to be conclusive. These are Cra\,I, Eri\,II, Gru\,I, Hya\,II, Leo\,I, Leo\,II, Leo\,IV, Phx\,I, and Pis\,II. When the orbital pole directions are Monte Carlo sampled from the measured uncertainties, there about a one in three chance that the pole is aligned with the VPOS ($0.13 < p_{\mathrm{inVPOS}} < 0.43$). These are thus consistent with aligning with the VPOS within their uncertainties. Even if their poles were intrinsically as well aligned as geometrically possible, due to the considerable uncertainty in their orbital pole directions their poles are expected to be found outside of  $\theta_{\mathrm{inVPOS}}$\ with probabilities of $p_{\mathrm{>VPOS} }= 37$\ to 60 per cent. With $p_{\mathrm(>obs)} = 10$ to 51 per cent it is expected that the poles are as far or further from the VPOS as observed, except for Leo\,I and Gru\,I for which these probabilities are only 1.1 and 0.4 per cent, respectively.  Three of the seven are more likely counter-orbiting, the remaining five more likely co-orbiting relative to the VPOS, so the counter-orbiting fraction is $f_\mathrm{counter} = \frac{3}{7} = 0.43$.

Of the remaining seven satellites which have orbital poles that are misaligned with the VPOS normal, one is consistent with aligning to within $\theta_{\mathrm{inVPOS}}$\ at a 6 per cent level (Boo\,I). However, it is basically impossible that this object align as well as geometrically possible with the VPOS, because $p_{\mathrm{>VPOS}} \approx 0$. This means that if this satellite orbits in the VPOS, then its pole can not be quite as close to the VPOS normal as would be geometrically possible. The satellite is most-likely counter-orbiting.

Thus, only six of the satellites have orbital poles that are firmly and conclusively misaligned with the VPOS even though they could have had aligned poles. These are Boo\,II, Car\,II, CBer\,I, Sxt\,I, Tuc\,II, and UMa\,I. They all have well constrained orbital pole directions $p_{\mathrm{inVPOS}} \leq 0.02$\ and  $p_{\mathrm{>obs}} = 0$\ thus are not orbiting along the plane defined by the VPOS. Three of the six are most-likely counter-orbiting ($f_\mathrm{counter} = 0.5$). For a more in-depth analysis in regard to the VPOS, we refer the reader to Pawlowski et al. (in prep.).

   \begin{table}
      \caption[]{Alignment with the VPOS. Column 1 gives the name of the satellite, col. 2 the angle between the predicted orbital pole and the VPOS normal, col. 3 the angle between the median measured orbital pole and the VPOS normal, col. 4 the fraction of Monte Carlo realizations that have an orbital pole aligned with the VPOS to within the 10 per cent circles, col. 5 the probability to falsely find an intrinsically perfectly aligned orbital pole outside of this area given the measurement uncertainties, and col. 6 the probability to find an orbital pole at least as far inclined from the VPOS as the median measured orbital pole.}
         \label{VPOS2}
     $$
         \begin{array}{p{0.1\linewidth}lllll}
            \hline
          
              \hline
Name        & \theta^{\mathrm{pred}} [^\circ] & \theta^{\mathrm{obs}} [^\circ] & p_{\mathrm(inVPOS)} & p_{\mathrm(>VPOS)} & p_{\mathrm(>obs)} \\ \hline
AquII       &    7.8  &  -19.6  &  0.532  &  0.449  &  0.709  \\
BooI        &   16.1  &  -50.5  &  0.057  &  0.004  &  0.000  \\
BooII       &   12.8  &  -55.8  &  0.014  &  0.001  &  0.000  \\
CVenI       &    1.5  &   24.9  &  0.381  &  0.473  &  0.609  \\
CVenII      &    4.2  &   -7.0  &  0.597  &  0.404  &  0.889  \\
CarI        &    4.7  &    8.0  &  1.000  &  0.000  &  0.354  \\
CarII       &    3.5  &   43.1  &  0.013  &  0.000  &  0.000  \\
CarIII      &    7.1  &    9.8  &  1.000  &  0.000  &  0.015  \\
CBerI       &    9.6  &   82.0  &  0.000  &  0.000  &  0.000  \\
CraI        &    9.6  &   40.9  &  0.335  &  0.489  &  0.436  \\
CraII       &   15.2  &   26.4  &  0.660  &  0.176  &  0.390  \\
DraI        &   10.4  &   18.3  &  0.999  &  0.000  &  0.029  \\
DraII       &   29.9  &   36.9  &  0.479  &  0.000  &  0.000  \\
EriII       &    9.3  &  -69.8  &  0.189  &  0.372  &  0.104  \\
FnxI        &   14.6  &   28.2  &  0.874  &  0.003  &  0.026  \\
GruI        &   24.9  &  -89.0  &  0.128  &  0.459  &  0.004  \\
HerI        &   37.7  &  -37.8  &  0.000  &  1.000  &  0.944  \\
HorI        &    1.0  &    4.9  &  0.994  &  0.002  &  0.676  \\
HyaII       &   28.9  &   63.7  &  0.145  &  0.600  &  0.199  \\
HyiI        &   10.5  &   17.2  &  1.000  &  0.000  &  0.000  \\
LeoI        &   20.4  &   88.3  &  0.229  &  0.490  &  0.011  \\
LeoII       &   13.4  &   50.4  &  0.388  &  0.556  &  0.393  \\
LeoIV       &    2.7  &  -58.1  &  0.395  &  0.444  &  0.237  \\
LeoV        &    1.1  &  -19.3  &  0.401  &  0.621  &  0.795  \\
PhxI        &    0.8  &  -40.7  &  0.427  &  0.502  &  0.454  \\
PisII       &    4.6  &   43.9  &  0.330  &  0.580  &  0.515  \\
RetII       &   11.9  &   13.4  &  1.000  &  0.000  &  0.020  \\
SgrI        &   60.2  &  -76.1  &  0.000  &  1.000  &  0.000  \\
SclI        &    5.0  &   -7.3  &  1.000  &  0.000  &  0.247  \\
Seg1        &   35.1  &  -35.3  &  0.640  &  0.179  &  1.000  \\
Seg2        &   58.7  &  -77.9  &  0.000  &  1.000  &  0.000  \\
SxtI        &   14.7  &   66.7  &  0.000  &  0.000  &  0.000  \\
TriII       &   64.5  &  -65.1  &  0.000  &  1.000  &  0.094  \\
TucII       &   25.3  &   49.7  &  0.001  &  0.004  &  0.000  \\
TucIII      &    6.1  &  -16.6  &  0.827  &  0.031  &  0.298  \\
UMaI        &   36.0  &  -55.1  &  0.001  &  0.555  &  0.003  \\
UMaII       &   55.4  &  -74.8  &  0.000  &  1.000  &  0.000  \\
UMiI        &   21.7  &   29.4  &  0.899  &  0.000  &  0.007  \\
Wil1        &   39.1  &   83.7  &  0.000  &  1.000  &  0.003  \\
                   \noalign{\smallskip}
            \hline
         \end{array}
     $$
   \end{table}

   \subsection{Group Infall}
   \label{Group-infall}
   
 The orbital pole of the LMC is similar to the direction of
the VPOS normal. Therefore our proper motion
measurements suggest that several of the dwarf galaxies in the sample
that have orbital poles falling within 10\% of
the sky around the assumed VPOS normal in the co-rotating assumption might
be associated to the LMC. The association of
MW dwarf galaxies to the Magellanic system is the subject of other works \citep{Kallivayalil_18,Fritz_18b}. 

Mainly on the basis of the objects position on the sky and 
heliocentric distance, \citet{Torrealba_16b} argued that Crater~I, Crater~II, Leo~II, Leo~IV and Leo~V might have 
once formed part of a group accreted by the MW. Even though the errors on the orbital pole determinations are large,
our analysis suggests such a prior physical association of all five objects together is unlikely. Given the large measurement uncertainties, the poles of Leo~IV and V are rather unconstrained, but compatible with each other.  An association of Cra~I and Cra~II appears likely, since several of the MC realizations bring Cra~I orbital pole in agreement to the region occupied by Cra~II pole, and the two systems share very similar (or compatible within 1$\sigma$) orbital properties.

In contrast the poles of Cra~II and Leo~II, which are are better constrained, do not overlap well. The preferred pericenter of these two objects differs by about 30\,kpc , but the values agree  within 1\,$\sigma$, therefore this cannot be used as a discriminator; also the orbital properties of Cra~I are compatible with those of Leo~II.

Thus, we cannot exclude the possibility that there are two groups, one 
including Cra~I and Cra~II, and the other one with  Leo~IV and Leo~V, with Leo~II being possibly part of one group or the other one; but we deem it unlikely 
that all the five objects came together as one group. Since, however the HST proper motion measurements places Leo~II's orbital pole in  the VPOS direction \citep{2013ApJ...768..139S, 2016AJ....152..166P, Pawlowski_2017}, it might be possible hat none of the measurement of Leo~II is accurate enough and that further studies with either HST or \textit{Gaia} are necessary.

\section{Summary and conclusions}
We derive systemic proper motions for all dwarf galaxies or galaxy candidates within 420\,kpc using 
\textit{Gaia} DR2 proper motions, for which literature spectroscopic members are available. 
Our proper motion determinations are in very good agreement (usually within 1$\sigma$) with the determinations by \citet{Helmi_18} and \citet{Simon_18,Massari_18,
Pace_18}. 

We derive the implied Galactocentric
velocities, and calculate orbits in canonical MW halo potentials of
"low" ($0.8 \times 10^{12} M_\odot$) and "high" mass ($1.6 \times
10^{12} M_\odot$). Comparison of the distributions of orbital apocenters
and 3D velocities to the halo virial radius and escape velocity,
respectively, suggests that the satellite kinematics are best explained
in the high-mass halo.  Relevant to the missing satellite problem, the fact
that fewer galaxies are observed to be near apocenter than near
pericenter implies that there must be a population of distant dwarf
galaxies yet to be discovered.

Several satellites have likely pericenter distances reaching within 20\,kpc from the Milky Way centers, and are therefore candidates for having suffered strong tidal disturbance (Tuc~III, Sgr~I, Her~I, Dra~II , Cra~II, Tri~II, Seg~1 and Wil~1). Among these, the orbital properties of Tuc~III, Sgr~I, Her~I, Dra~II and the ``feeble giant'' Crater~II are in line with the detections of tidal features in these objects and the predictions of models. In contrast, no tidal features have been detected yet in Tri~II, Seg~1 and Wil~1 . This difference could be partly caused by the still relevant errors in the derived pericenter values and/or by the intrinsic difficulty in detecting tidal features, especially in very faint systems.  Our analysis suggests that  also a couple of classical MW dSphs, such as Draco~I and UMi, have internal orbits with pericenter distances that bring them dangerously close to the internal regions of our Galaxy, at risk of being  tidally affected. It appears then crucial to address in more detail the impact that tidal disturbances 
might have had in the structural and internal kinematic properties of these galaxies.

Of the 23 satellites for which we can draw conclusions, 17 are orbiting along the plane of satellites (VPOS, not counting the well aligned LMC and SMC) and 6 are not. These findings suggest that a majority of the MW satellites for which we have measured proper motions orbits along the VPOS, but that not all satellites participate in coherent motion along this structure. This is in line with an analysis based solely on the spatial distribution of satellite galaxies, which found that up to half of the MW satellites might be drawn from an isotropic distribution in addition to satellites drawn from a planar distribution \citep{2016MNRAS.456..448P}, as well as with the satellite plane around Andromeda which consists of about half of the M31 satellites \citep{2013Natur.493...62I}.

The distribution of orbital poles does not appear to confirm the hypothesis that Crater~1, Crater~II, Leo~II, Leo~IV and 
Leo~V were all accreted by the MW as part of the same galaxy group; although the errors are large and we cannot exclude 
an association between Crater~1 \& Crater~II or Leo~II with Leo~IV \& Leo~V, it seems unlikely that these objects are 
all associated to each other.

Finally, we note that we just use part of the power of \textit{Gaia} DR2 for the determination of systemic proper motions of dwarf galaxies and 
this has already led to constraining proper motions for dozens of galaxies. We expect that in some cases adding stars without 
existing spectroscopic measurements should improve the precision. Since the precision in proper motion determinations grows with the 1.5 power of the time-baseline (and systematics also, when they are not based on a moving reference frame) we expect that the proper motions of \textit{Gaia} should be a factor 4.5 more accurate after the nominal mission and possibly a factor 12 after the extended mission. This would enable to measure 
systemic proper motions for essentially all galaxies in the sample with the best precision now possible for very few galaxies. For example, 
even the motion of Phoenix~I could be measured to about 
19 km/s precision, similar to the third best measurement of HST to date. \\

\begin{acknowledgements}
This work has made use of data from the European Space Agency (ESA) mission
{\it Gaia} (\url{https://www.cosmos.esa.int/gaia}), processed by the {\it Gaia}
Data Processing and Analysis Consortium (DPAC,
\url{https://www.cosmos.esa.int/web/gaia/dpac/consortium}). Funding for the DPAC
has been provided by national institutions, in particular the institutions
participating in the {\it Gaia} Multilateral Agreement.
      
G.B. gratefully acknowledges financial support by the Spanish Ministry of Economy and Competitiveness (MINECO) under the Ramon y Cajal Programme (RYC-2012-11537) and the grant AYA2014-56795-P. 
MSP acknowledges that support for this work was provided by NASA through Hubble Fellowship Grant \#HST-HF2-51379.001-A awarded by the Space Telescope Science Institute, which is operated by the Association of Universities for Research in Astronomy, Inc., for NASA, under contract NAS5-26555. NK is supported by the NSF CAREER award 1455260.
This project is also part of the HSTPROMO (High-resolution Space Telescope PROper MOtion) 
Collaboration\footnote{http://www.stsci.edu/~marel/hstpromo.html}, a set of projects aimed at improving our dynamical understanding of stars, clusters and galaxies in the nearby Universe through measurement and interpretation of proper motions from HST, \textit{Gaia}, and other space observatories. We thank the collaboration members for the sharing of their ideas and software.
We thank Josh Simon, Nicolas Martin and Matthew Walker for providing the catalogs of \citet{Simon_07}, \citet{Martin_16} and an improved catalog of \citet{Walker_16}, respectively.
We thank Matteo Monelli for guidance on the bibliography for
distance estimates from variable stars.

\end{acknowledgements}

\begin{appendix} 
\section{Details on galaxies}
\label{ap_det_galaxies}

In this Appendix we first give details on aspects concerning all or several of the objects in the sample, and then provide specific details for each object. 

$\bullet$ For the systematic error in proper motion, we need to estimate the size of each galaxy or candidate galaxy in the sample. We determine this by taking the average extent of member stars in R.A. (corrected by $\cos(\mathrm{Dec.}$)) and Dec. and averaging between these two values. For example, for Aqu~II this leads to a size of 0.019 degree and thus a systematic error of 0.063 mas/yr. 

$\bullet$ For spectroscopic catalogs not containing membership information, we adopt different selection criteria, depending on whether only line-of-sight velocities are available from the literature, or also other additional information that allows to classify a star as a giant, such as the star's surface gravity, log$g$, or the gravity-sensitive criterion in  \citet{BattagliaSt_12}, based on the equivalent width (EW) of the NIR Mg~I line as a function of the EW of the near-IR Ca~II triplet.  The gravity or gravity-sensitive criterion are applied only to systems for which the spectroscopic members are indeed expected to be giant stars. 

When only line-of-sight velocities are used, we assign probabilities  ($p$) with the following equation:
$$p=\exp(-1/a*(v_\textrm{LOS}- v_\textrm{LOS,sys})^2).$$
while when also log$g$ estimates are available:
$$p=(\tanh([c-log(g)]/b)+1)/2*\exp(-1/a*(v_\textrm{LOS}- v_\textrm{LOS,sys})^2)$$
where $v_{\rm LOS}$ is the heliocentric line-of-sight velocity of a given star, $v_{\rm LOS,sys}$ the systemic velocity in the same system; $a$, $b$ and $c$ are normalization constants that depend on the system and are chosen to yield a probability of $\sim$0.5 at heliocentric velocities about 3x the internal l.o.s. velocity dispersion of the system, as given in the source papers.

For Fornax and Sculptor, when considering the \citet{BattagliaSt_12} catalogs, we adopt their same selection of  likely giant stars. 
\subsection{Aquarius~II}
For Aquarius~II just the two brightest members (according to the binary classification) of \citet{Torrealba_16a} have matches in GDR2.
However, we consider its systemic motion reliable because the star density in the stellar halo is low at Aquarius' distance (105 kpc) and because 3 stars with similar colors without spectroscopy have a consistent proper motion, see also \citet{Kallivayalil_18}.

\subsection{Bo\"{o}tes~I}
We use two spectroscopic catalogs for this galaxy, from \citet{Martin_07} and \citet{Koposov_11}: the former contains binary membership
classification and we adopt from that study also the values of the heliocentric l.o.s. systemic velocity ($v_\textrm{LOS,sys}= 99$km/s) and
velocity dispersion (5 km/s); no membership classification is given in the latter study.

For stars that were observed in both works, we adopt the binary classification of \citet{Martin_07}. To those that
are only in the catalog of \citet{Koposov_11}, we assign probabilities  on the basis of the l.o.s. velocity. 
Our selection is somewhat different from that of \citet{Simon_18} but we agree within one\,$\sigma$ in the systemic proper motion.  
We also agree within 1\,$\sigma$ with the PM determination by \citet{Helmi_18}, who used a selection based on photometry.

\subsection{Bo\"{o}tes~II}

We find matches in GDR2 for 4 bright members (according to the binary classification) in \citet{Koch_09}. We consider Bo\"{o}tes~II systemic PM reliable
since also some other stars whose colors are compatible with Boo~II share the same motion. Our determination agrees within 1\,$\sigma$ with
that by \citet{Simon_18}. 

\subsection{Canes Venatici~I}
We use the catalogs of \citet{Martin_07} and \citet{Simon_07} and the  binary probabilities there-in. We give the classifications of \citet{Martin_07} higher priority. Since 57 stars pass all our cuts, the derived motion is robust.
\subsection{Canes Venatici~II}
We use the catalog of \citet{Simon_07}. From its binary probabilities 11 members remain after all cuts. 

\subsection{Carina I}
We use the catalog of \citet{Walker_09_clas} with continuous  probabilities. 
Our proper motion agrees within 1\,$\sigma$ with that by \citet{Helmi_18}. For both works the systematic error dominates over the statistical uncertainty. 
\subsection{Carina II}
We use the catalog of \citet{Li_18a} which has binary probabilities.
The 18 member stars we select form a clear clump in proper motion space supported by stars w/o spectroscopy. Our motion agrees within 1\,$\sigma$ with \citet{Simon_18}, \citet{Kallivayalil_18} and \citet{Massari_18}.

\subsection{Carina III}
Again we use the catalog of \citet{Li_18a} which has binary probabilities. The 4 member stars we select form a clear clump in proper motion space supported by one bright star w/o spectroscopy. Our motion agrees within 1\,$\sigma$ with \citet{Simon_18} and \citet{Kallivayalil_18}.
\subsection{Coma Berenices I}
We use the catalog of \citet{Simon_07} with its binary membership classifications. After all cuts this leads to 17 members in our final selection. They form a clear clump in proper motion space supported by a few stars without spectroscopic information.  Our proper motion agrees  within 1\,$\sigma$ with the motion of \citet{Simon_18}.

\subsection{Crater~I}
For this system we give higher priority to the continuous  probabilities by \citet{Voggel_16}. If that is not available we use the binary classification of \citet{Kirby_15}. Ten members stars remain after the membership selection. Due to Crater\,I relative high surface brightness, pollution by contaminant stars in not on issue. Because the available spectroscopy is deep and rather complete there are no additional member candidates with \textit{Gaia} DR2 information only. 

\subsection{Crater~II}

In this case,  probabilities are not included in the tables of \citet{Caldwell_17}, therefore we calculate them using both l.o.s. velocities (with a systemic velocity of 87.5 km/s) and log$g$ .
For log(g) our selection is guided by Figure~3 of \citet{Caldwell_17}. 
With our selection, 59 stars have a probability greater than 0.4. That is slightly less than in \citet{Caldwell_17}, where 62 stars have a probability greater than 0.5, but on the safe, conservative side.   If we were to change the probability cut to 75\%, this would yield a  0.56 $\sigma$ difference in proper motion for the declination component. This is 0.019 mas/yr, less than the systematic error, therefore we consider our motion as robust. 

Recently, \citet{Kallivayalil_18} added 59 stars to our spectroscopic sample, selecting them from photometry, and obtained a 4\,$\sigma$ different proper motion. Most of the added stars are rather faint, thus it is not clear that adding them improves the overall accuracy, since it is difficult to be sure that these faint stars are truly members, especially since Crater~II has a really low surface brightness. It also could be that assignment of binary membership is not sufficient for photometric members, and that a probabilistic treatment like in \citet{Pace_18} is advisable. Also there seem be a problem for Tuc~II with the motion of \citet{Kallivayalil_18}, see Section~\ref{subsec:TucII}.
Therefore, we prefer the proper motion derived here.

\subsection{Draco~I}
We use the catalog of \citet{Walker_15}. This does contain probabilities and we calculate them using the l.o.s. velocities and star's gravity. 
Our motion agrees within 1\,$\sigma$ with \citet{Helmi_18}

\subsection{Draco~II}
We use the catalog of \citet{Martin_16} and find 6 stars matching all our criteria. The derived PM agrees very well with that by  \citet{Simon_18}. 
There are  a few faint stars w/o spectra whose motion agrees with the spectroscopic selected members \citep{Kallivayalil_18}.

\subsection{Eridanus II}
We use the catalog by \citet{Li_17} with binary probabilities, 12 of which are  bright enough to have \textit{Gaia} DR2 kinematic measurements. Due to the galaxy faintness and large distance, the errors on the PM are large.

\subsection{Fornax (I)}

We use the catalogs of \citet{Walker_09_clas,BattagliaSt_12}.
For both data sets we use the LOS velocities to select members for the \citet{BattagliaSt_12} data set also the gravity indicators.
Due to the brightness of that galaxy, our proper motion does not depend on the membership probability cut adopted. Our proper motion differs in Dec. by about 2 $\sigma$ (our statistical error), or 0.012 mas/yr from \citet{Helmi_18}. Since the spatial coverage of the data we use differs from that in the \textit{Gaia} collaboration analysis, this discrepancy  could be caused by spatial variations in the systematic error or by real physical differences over the  angular area covered by this, rather large, galaxy. 

\subsection{Grus~I}
For Grus~I we use the catalog by \citet{Walker_16}. We note that the probabilities in the long electronic Tab.~1 of \citet{Walker_16} are incorrect for Grus~I. We use a corrected version kindly provided by the lead author. The probabilities listed in Tab.~1 in the pdf version of that article are correct and contain all members besides one rather faint star. 
Grus~I is one of the few satellites whose systemic PM is somewhat sensitive to which stars are included/excluded. The two brightest stars of its 5 likely members have with $p=0.69$ (Gru1-032) and $p=0.68$ (Gru1-038), i.e. relatively low membership probabilities. Excluding those changes the velocity by 0.7/0.9\,$\sigma$ in the two dimensions. We decided to include these stars since both their PMs are within our proper motion selection box even without considering their error-bars.  Our systemic PM agrees with that in \citet{Kallivayalil_18}, which is unsurprising,  since we mostly use the same stars. A motion also agrees with the purely photometric proper motion determination of \citet{Pace_18}.

\subsection{Hercules (I)}
In case of Hercules we assign the highest priority to the work of \citet{Simon_07}.  We complement the analysis with the stars in \citet{Aden_09}, which  are were all classified as members in that study, even though not all of them have spectroscopic information, e.g. as some candidate horizontal branch. Since the stars without spectroscopy are faint, excluding them does cause significant changes in the average velocity, given the error-bars.   We therefore  include them in our final estimate.

\subsection{Horologium I}
For Horologium I we use the catalog of \citet{Koposov_15b} which contains binary membership classification. We remain with 4 members after all our cuts. These relatively bright stars form a tight group in proper motion space. 
Our proper motion agrees very well with that in \citet{Simon_18}, \citet{Kallivayalil_18} and \citet{Pace_18}.

\subsection{Hydra II}
We use the catalog of \citet{Kirby_15}, which contains binary membership classification. After our cuts we have 6 members, whose motion agree well with each other. Our systemic PM agrees also very well with the determination by \citet{Kallivayalil_18}.

\subsection{Hydrus~I}
In the case of Hydrus~I a spectroscopic catalog with continuous membership probabilities is published in
the electronic table of \citet{Koposov_18}. Of the 33 candidates, 30 remain after our cuts. One remaining member has a relatively low membership probability of 56\%; if excluded, the motion changes by 0.1/0.4 $\sigma$ in R.A./Dec.
\citet{Kallivayalil_18} added some more photometric candidate members. Our motion agrees with their motion and also with \citet{Simon_18} within about 1\,$\sigma$.

\subsection{Leo~I}

For Leo~I we use the catalog of \citet{Mateo_08}, which does not contain membership probabilities nor gravities. Therefore we assign a probability on the basis of the l.o.s. heliocentric velocity.
 After our cuts we have 241 members and thus the systemic motion is robust. We find good agreement with the  \citet{Helmi_18} determination.

\subsection{Leo~II}

For this galaxy we use primarily the catalog of \citet{Spencer_17}, which has binary probabilities. In addition, we complement the analysis with the catalog of \citet{Koch_09}, to which stars we assign membership probabilities on the basis of l.o.s. heliocentric velocities. 
Combining both catalogs, 131 members pass all our cuts.
Our motion agrees well with that of \citet{Helmi_18}.

\subsection{Leo~IV}
For this galaxy we use the catalog of \citet{Simon_07}, which provides binary probabilities. Due to the galaxy distance and faintness, only 3 members remain after our cuts.
A few faint stars w/o spectroscopy have a consistent motion to our measurement. Also the background is low at the large distance of this galaxy. Due to the faintness of the three stars the error of the motion is still rather large and does not constrain much the properties of the system.

\subsection{Leo~V}
We use the catalog of \citet{Walker_09} with continuous membership probabilities. After our cuts we have 5 member stars, all with a high spectroscopic membership probability, which should imply they are robustly classified as members.
In addition to the stars with spectroscopy, the is one faint star w/o spectroscopy with a consistent proper motion.
Due to the distance and faintness of the system, the error on the systemic PM is rather large and it is not very constraining of the orbital properties of the galaxy.

\subsection{Phoenix~I}
We use the catalog of \citet{Kacharov_17} which provides binary probabilities. Because the galaxy is rather bright we have relatively many members (71) even if the system is very distant. There are also more stars w/o spectroscopy that have proper motion consistent with our determination. The spectroscopy based motion is still not very constraining of the properties of the galaxy.

\subsection{Pisces~II}

We use the catalog of \citet{Kirby_15}, which provides binary probabilities. Only two stars (9004 and 12924 in the nomenclature of table 2 of \citet{Kirby_15}) pass our cuts and there are no additional stars w/o spectroscopy that we can use to back up our determination.  This is partly expected due to the faintness of the galaxy. Clearly, the systemic PM is less reliable than most of our determination, but its error is anyway rather large.

\subsection{Reticulum~II}
We use the catalog of \citet{Simon_15}, giving binary membership classifications. After our cuts we have 27 member stars. Our motions agrees very well with \citet{Simon_18}, \citet{Kallivayalil_18} and \citet{Pace_18}. With agrees within 1.7/0.3\,$\sigma$ with the motion of \citet{Massari_18}.

\subsection{Sagittarius~I}
For the Sagittarius dSph we use APOGEE DR14 data  \citep{Majewski_17}, selecting all the stars within 30' of the nominal galaxy center. To assign membership probabilities we use a selection based on heliocentric l.o.s. velocity and stellar gravity, but retaining 
only very bright giants, because due to Sagittarius location close to the bulge and due to the selection function of APOGEE, also not member giants were observed. 
Sag~I is one of the galaxies were the systemic PM changes more noticeably (by -1/-0.6\,$\sigma$ in R.A. and Dec) if we were to adopt a stricter membership cut; nonetheless, this is just a change of 0.01 mas/yr or less.
Our systemic PM differs from the determination in \citet{Helmi_18} by 5\,$\sigma$ (0.044 ms/yr in $\mu_{\alpha}$). However, the difference is not much larger than the systematic error, which it could play a role for this large galaxy. It is likely that the fact that Sag~I is rather close also causes measurable proper motion gradients throughout the object, which could be present in varying amounts in samples with different spatial distributions.  In order to achieve a higher accuracy, it might be necessary  to consider a model accounting for possible such gradients.

\subsection{Sculptor~(I)}
As for Fornax, we use the catalogs of \citet{Walker_09_clas,BattagliaSt_12} and the same approach to member selection. We find good agreement with the motion of \citet{Helmi_18}.

\subsection{Segue~1}
We use the catalog of \citet{Simon_11} and the membership probabilities based on a Bayesian approach. Of the selected 13 stars, the star with the lowest  probability  has a $p =$ 0.885, thus the adopted cut does not matter. The same is true if we use one of the other two sets of probabilities provided in the article. 
Our motion agrees within about 1 $\sigma$ with \citet{Simon_18}. It is different by 2.2/0.2\,$\sigma$ (in R.A./Dec.) from the proper motion measurement of \citet{Fritz_18}), based on LBT data.

\subsection{Segue~2}
We use only the stars classified as certain members in \citet{Kirby_13}.  After our cuts,  we are left with 10 stars, which however do not  form a clear clump in proper motion space. 
Our systemic PM agrees in R.A. with the motion of \citet{Simon_18} but differs by 2\,$\sigma$ in Dec, while it differs by 3.1/3\,$\sigma$ (in R.A./Dec.) from the determination of \citet{Massari_18}. In contrast to us, the latter work includes also stars w/o spectroscopy. While adding these additional stars improves precision, it is not clear whether that is also true when uncertain membership is taken into account. We note that there is a stream at a similar distance as Segue~2 \citep{Belokurov_09,Kirby_13}, and that \citet{Massari_18} use of projected distance from the galaxy center as a criterion to start isolating targets is based on a simple cut. 
Because of Segue~2 close distance, pollution by other halo components is more important for this object than for most of the other systems in our sample. For an improved determination of the systemic PM, it might be necessary to wait for future \textit{Gaia} data releases in which the proper motion errors are smaller and thus the member stars would form a tighter clump.

\subsection{Sextans~(I)}
We use the catalog of \citet{Cicuendez_18} with continuous probabilities. With our standard cuts we select 325 members. The proper motion changes by less than 0.1\,$\sigma$ when we change the spectroscopic probability cut to P$>$0.7.  Our motion differs by 1.6/0.7\,$\sigma$ (in R.A./Dec.) from the motion of \citet{Helmi_18}, which might happen by chance or due to additional systematic differences due to the large angular size of this galaxy.

\subsection{Triangulum~II} 
We use the catalog of \citet{Kirby_17} which provides binary memberships. After our cuts we are left with 5 members; the star classified as uncertain member in the source paper is eliminated by our cuts.
These stars do not form a clear clump in proper motion space, but due to the faintness of the stars, and the possible larger population due to Tri~II low Galactic latitude, a large scatter is not surprising. Our motion agrees very well with the motion of \citet{Simon_18}.

\subsection{Tucana~II} 
\label{subsec:TucII}

We use the catalog of \citet{Walker_16}. It provides non binary memberships. With our standard cuts we obtain 19 members stars. Most of those form a clump in proper motion space. Some of the others are more scattered, but since they are fainter they not matter much for the overall motion. Our motion agrees within 1 $\sigma$ with the determination of \citet{Simon_18}, but differs by 0.9/3.0\,$\sigma$ in R.A./Dec with the determination by \citet{Kallivayalil_18}. Our agrees nearly perfectly with \citet{Pace_18}. This is evidence that the method of \citet{Kallivayalil_18} of adding photometric members is less reliable than the photometric method of \citet{Pace_18}.

\subsection{Tucana~III} 
This satellite has a short tidal stream \citep{Drlica-Wagner_16} that was covered spectroscopically by \citet{Li_18b}. Along streams of globular clusters (we note that Tuc~III has at most a dispersion of 2 km/s \citep{Li_18b,Simon_17}, thus is dynamically very similar to a globular cluster), gradients in proper motions are smooth and show only a small jump around the main body; for Pal~5, which is closer in distance than Tuc~III, this is less than 0.1 mas/yr. Such contribution would be of the same order as systematic uncertainties for \textit{Gaia} \citep{Lindegren_18} over small areas, therefore we decided to include stream stars. 
To avoid that a somewhat uneven coverage of both sides of the stream could have some influence on the results, we fit the the proper motion as function of position and use here only the proper motion at the intercept, i.e. at the position of Tuc~III. 
The systemic proper motion so obtained is consistent with the value in \citet{Simon_18}, who do not use stream stars and with \citet{Pace_18} who use less stream stars than we. Our results is also consistent with \citet{Kallivayalil_18} who obtain the proper motion in a more simple way. 

\subsection{Ursa Major I} 
We use the catalog of \citet{Martin_07} and, for stars not contained in that work, the catalog of \citet{Simon_07}. Both provide binary classifications. After our cuts we have 23 members. Our motion agrees very well with the determination of \citet{Simon_18}.

\subsection{Ursa Major II} 
We use the catalogs of \citet{Martin_07} and, for stars not contained in that work, the catalog of \citet{Simon_07}. Both provide binary classifications. After our cuts we have 15 members. These stars form a clear clump in proper motion space supported by some stars w/o spectroscopy. Also in this case our motion agrees very well with the determination of \citet{Simon_18}.

\subsection{Ursa Minor (I)} 
We use the catalog of \citet{Kirby_10}. It supposedly only contains members. After our cuts we have 137 stars. The systemic PM agrees very well with the determination by \citet{Helmi_18}.

\subsection{Willman 1} 
We use the catalog of \citet{Martin_07} which provides binary memberships. After our cuts we have 7 member stars. The brightest of those form a group in proper motion space which is supported by one star w/o spectroscopy. Our motion agrees within 1 $\sigma$ with the determination of \citet{Simon_18} but our error is relatively large.

\section{Monte Carlo simulations for errors}
\label{mc_errors}

\subsection{Forward MC simulations}
\label{mc_forward}
We run 2000 Monte Carlo simulations drawing 
$\mu_{\alpha^*}$, $\mu_{\delta}$ and the heliocentric l.o.s. velocities 
randomly, from Gaussian distributions centered 
on the measured values and dispersions given by the respective errors. When extracting random values for 
$\mu_{\delta}$ we do consider the correlation between the $\mu_{\alpha^*}$ and $\mu_{\delta}$, 
listed in Table~\ref{KapSou2}. We also use the listed value for the systematic error. 
The randomly simulated kinematic and positional properties 
are then transformed into velocities in a Cartesian heliocentric (and then Galactocentric) reference system. From these we then calculate orbital poles.
As we tested in Monte Carlo simulations, the derived quantities mentioned above are not biased, since they can take both positive and negative values. 

\subsection{Backward MC simulations}
\label{mc-back}

As shown in Fig.~\ref{mc-back}, the process of normal "forward MC simulations" described above obtains biased errors and values for positive defined properties like $v_\mathrm{tan}$ and $v_\mathrm{3D}$ when errors are close to be as large or larger than the observed value; this is because the error 
already affects the observed value and the standard MC simulations would add the error a second time. 
Therefore we adopt a different approach for estimating $v_\mathrm{tan}$ and $v_\mathrm{3D}$, and the corresponding errors.  Essentially, we want to determine what combinations of proper motions and l.o.s. velocities would result into the observed values of $v_{\rm tan}$ and $v_{\rm 3D}$ when they are convolved with their observational errors (see also \citet{Marel_08} for an independent development of a similar
methodology).  To this end we proceed in the following way. We draw values of $\mu_\mathrm{tot}$ (that is $\sqrt{\mu_{\alpha,*}^2+\mu_{\delta}^2}$) from a large grid of values, assuming a uniform distribution in $\mu_\mathrm{tot}$ and in its direction. In addition, we draw most of the other properties (e.g. l.o.s. velocity) as previously, with Gaussian errors. We then add the observed errors to $\mu_{\alpha,*}$ and $\mu_{\delta}$ and convert the proper motions and l.o.s. heliocentric velocities into  $v_\mathrm{tan}$ and  $v_\mathrm{rad}$. 
At this point, we require that both velocities agree with their corresponding observed  values within 0.05 to 0.8\,$\sigma$. We choose a tolerance of these sizes because for larger tolerances the retrieved values are bigger because numbers are somewhat more likely drawn on the larger side of the measured value. Our tolerance is larger for cases which have a more significant motion because biases are smaller in this case. 
Note that here we neglect the error in distance modulus to avoid that the velocity selection selects also in distance. For the errors we use the errors obtained in the forward MC simulations. 
For all the galaxies we obtain at least 2000 accepted Monte Carlo values: this gives us a distribution of 2000 error-free proper motions that, when convolved with their errors (and associated to the l.o.s. velocity and distance of the galaxy), would produce unbiased tangential and 3D velocities. The distribution of 2000 simulated $v_\mathrm{rad}$, $v_\mathrm{tan}$ and $v_\mathrm{3D}$ is then simply derived by transforming the 2000 error-free proper motions and Gaussianly distribution l.o.s. velocities into Galactocentric velocities; in this step, we allow for distance errors. We use quantiles to give 1\,$\sigma$ errors in $v_\mathrm{rad}$, 
$v_\mathrm{tan}$ and $v_\mathrm{3D}$, and use as most likely value the median of the distribution. 
For $v_\mathrm{rad}$ the errors are nearly symmetric, thus we give a single error, for the other two we give asymmetric errors in Table~\ref{KapSou2}.

\begin{figure}
   \centering
   \includegraphics[width=0.72\columnwidth,angle=-90]{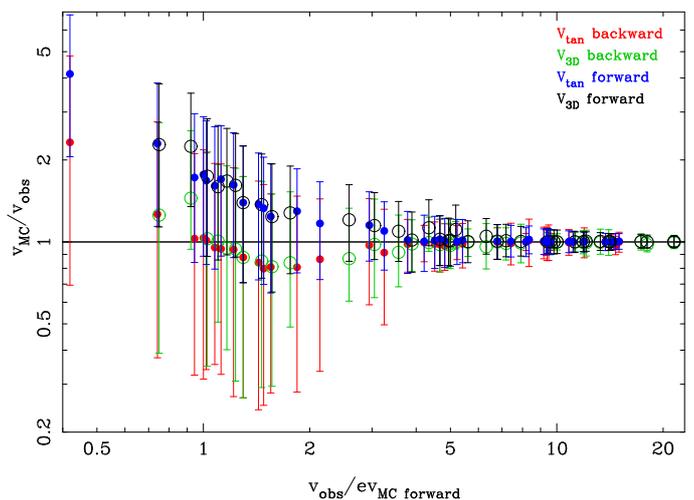}   
      \caption{Comparison of errors and values for positive-defined quantities such as $v_\mathrm{3D}$ and $v_\mathrm{vtan}$, using forward and backward Monte Carlo simulations. The x-axis gives the ratio of observed velocity over the error given by the forward MC simulations, while the y-axis shows the ratio of the median velocity obtained from the (forward or backward, see legend) MC simulations and the observed value.
              }
         \label{fig:mc-back}
   \end{figure}

We compare in Figure~\ref{fig:mc-back} the difference between forward and backward simulations. The outcome mainly depends on the error in proper motion, which is closely related to the error in tangential velocity; of smaller importance are the error in distance and line of sight velocity. It is visible that for $v_\mathrm{tan}$ of less than about 4\,$\sigma$ significance the tangential velocity is over estimated in the forward Monte Carlo simulations. The situation is similar for $v_\mathrm{3D}$ but, since in this quantity the line-of-sight velocity matters much more, the relationship is less linear. 
   
The bias in $v_\mathrm{tan}$ and $v_\mathrm{3D}$ also affects the estimate of orbital properties such as apocenter and pericenters, because galaxies with a 3D velocity that is artificially too large would appear to be less bound than they really are. Therefore we use the backward Monte Carlo simulations results 
(precisely
$\mu_{\alpha*}$, $\mu_\delta$, and $V_\mathrm{LOS}$) also in Galpy, selecting for each satellites a random subsample of 500 cases. We then retrieve the median and the one $\sigma$ range in the distribution of peri/-apocenter and eccentricity from the 500 orbital integrations. 

We note that because in the way the backward Monte Carlo simulations had to be set up, the orbit has a random orientation with respect to the disc, which can impact orbits  with a small pericenter. Tuc~III is the only satellite which is certainty affected; however since its pericenter is just 3 kpc, also other approximations, such as the use of a spherical bulge in Galpy, would matter, making the used gravitational potential anyway not fully realistic. All the other satellites have most likely pericenters of at least 14 kpc, such that the disk is less important than the positive bias of the forward Monte Carlo calculation.

\end{appendix}

\bibliographystyle{aa} 
\bibliography{mspap} 
\end{document}